\documentclass[conference,compsoc]{IEEEtran}

\usepackage{tikz}
\usetikzlibrary{arrows.meta}
\usepackage{amsmath}

\usepackage{filecontents}

\usepackage{algpseudocode}
\usepackage[linesnumbered, ruled,vlined]{algorithm2e}
\usepackage{algorithmicx}

\usepackage{tabularx}
\usepackage{hhline}
\usepackage{amssymb}%
\usepackage{pifont}%
\usepackage{booktabs} 
\usepackage{siunitx}

\usepackage{xurl}
\usepackage[hidelinks]{hyperref}

\usepackage{graphicx}
\usepackage{subcaption}

\usepackage{verbatim}
\usepackage[shortlabels]{enumitem}
\usepackage{cite}

\newcommand{\point}[1]{\par\smallskip{\noindent\textbf{#1.}~}}

\newcommand{\bolditem}[1]{\par\smallskip{\noindent\textbf{#1}~}}

\newtheorem{theorem}{Theorem}[section]
\newtheorem{lemma}[theorem]{Lemma}

\newcommand{\adversary}{$\mathcal{A}$\xspace}
\newcommand{\challenger}{$\mathcal{C}$\xspace}

\newcommand{\adversarymath}{\mathcal{A}\xspace}
\newcommand{\challengermath}{\mathcal{C}\xspace}

\newcommand{\continit}{$\textsc{ContactInit}$\xspace}
\newcommand{\lookup}{$\textsc{Lookup}$\xspace}
\newcommand{\register}{$\textsc{Register}$\xspace}
\newcommand{\addfriend}{$\textsc{AddFriend}$\xspace}

\newcommand{\unlinkabilitygame}{$\mathit{Exp}_{G1}$\xspace}
\newcommand{\impersonationgame}{$\mathit{Exp}_{G2}$\xspace}
\newcommand{\idverificationgame}{$\mathit{Exp}_{G3}$\xspace}
\newcommand{\unobservabilitygame}{$\mathit{Exp}_{G4}$\xspace}

\newcommand{\upperboundtotalmean}{13\,}
\newcommand{\registermean}{8.04\,}
\newcommand{\registerpninety}{10.02\,}
\newcommand{\registerupperboundmean}{8.5\,}
\newcommand{\lookupmean}{3.44\,}
\newcommand{\lookuppninety}{4.41\,}
\newcommand{\lookupupperboundmean}{4\,}
\newcommand{\anonmean}{8.21\,}
\newcommand{\anonpninety}{5.70\,}

\newcommand{\nonanonmean}{12.34\,}
\newcommand{\nonanonpninety}{11.86\,}
\newcommand{\nonanonupperboundmean}{13\,}

\begin{document}

\date{}

\title{Pudding: Private User Discovery in Anonymity Networks}

\author{
{\rm Ceren Kocaoğullar}\\
University of Cambridge
\and
{\rm Daniel Hugenroth}\\
University of Cambridge
\and
{\rm Martin Kleppmann}\\
TU Munich
\and
{\rm Alastair R. Beresford}\\
University of Cambridge
} %

\maketitle

\begin{abstract}
Anonymity networks allow messaging with metadata privacy, providing better privacy than popular encrypted messaging applications.
However, contacting a user on an anonymity network currently requires knowing their public key or similar high-entropy information, as these systems lack a privacy-preserving mechanism for contacting a user via a short, human-readable username.
Previous research suggests that this is a barrier to widespread adoption.

In this paper we propose Pudding, a novel private user discovery protocol that allows a user to be contacted on an anonymity network knowing only their email address.
Our protocol hides contact relationships between users, prevents impersonation, and conceals which usernames are registered on the network.
Pudding is Byzantine fault tolerant, remaining available and secure as long as less than one third of servers are crashed, unavailable, or malicious.
It can be deployed on Loopix and Nym without changes to the underlying anonymity network protocol, and it supports mobile devices with intermittent network connectivity. 
We demonstrate the practicality of Pudding with a prototype using the Nym anonymity network.
We also formally define the security and privacy goals of our protocol and conduct a thorough analysis to assess its compliance with these definitions.
\end{abstract}

\section{Introduction}\label{sec:intro}

As more of our conversations take place online, protecting them becomes increasingly important.
As a result, we have seen widespread deployment of end-to-end encryption in messaging apps in order to protect the confidentiality and integrity of our correspondence.
Today's secure messaging apps like Signal, iMessage, and WhatsApp allow users to contact their friends using only a low-entropy username, such as a phone number.
In contrast, earlier systems such as PGP required users to explicitly manage cryptographic key material, which was a significant usability issue and a barrier to adoption~\cite{whitten1999johnny}.
It is likely that user discovery via phone number played a large part in why Signal, iMessage, and WhatsApp succeeded in gaining widespread adoption, while PGP has remained a niche tool used by experts.

While the current generation of messaging apps offer confidentiality of message contents against malicious servers and network adversaries as long as the users verify each other's keys, these apps offer little \emph{metadata privacy}.
In particular, the service provider is able to track who is talking to whom, and at what times---information that can be sensitive \cite{ukrainecellphones, strohmeier2018real, fitbitheatmap, bellingcatnavalny}.
Even Signal's sealed sender mechanism \cite{signalsealedsender}, which aims to hide the sender of the messages from the service provider, does not provide protection at the network level.
In response, significant academic progress has been made in the design of practical anonymity networks to protect metadata privacy for messaging apps, such as Loopix \cite{loopix}, Vuvuzela \cite{van2015vuvuzela}, Talek \cite{talek}, and Pung \cite{pung}.
However, despite the fact that user discovery was important for widespread adoption of secure messaging, none of these networks consider user discovery as a part of their design.
Instead, they defer the problem to future work.
Yet, an automated means of supporting user discovery and key management is critical if we want to build usable messaging apps that reach a wide audience and support metadata privacy, as well as end-to-end encryption.
Such a system needs to allow a messaging app to securely and automatically resolve user-readable names entered in a contact address book (such as phone numbers or email addresses) into appropriate network addresses and the necessary cryptographic key material required to send encrypted messages.
Furthermore, such a system needs to do this while protecting metadata privacy.

The user discovery model found in the current generation of end-to-end encrypted messaging apps is unsuitable if we wish to provide metadata privacy, because this model undermines the metadata privacy that anonymity networks provide. 
Specifically, encrypted messaging apps such as Signal, iMessage, and WhatsApp store user public key material in trusted key directory servers, which users query with phone numbers to retrieve the associated public keys. 
This enables the servers to learn which contacts in a user's address book are registered to the messaging app (e.g. Signal \cite{signalcontactdiscovery}), and in some cases even those that are not registered (e.g. WhatsApp \cite{whatsapprivacypolicy}).
Communicating with a key directory via an anonymity network may provide metadata privacy against a network observer, but it does not provide protection against malicious or compromised directory servers. A compromised directory server may try to deceive a user into talking to someone other than the intended user.
Moreover, an adversary can learn the phone numbers of all registered users by querying the directory~\cite{hagen2021all}.

A relatively small body of literature is concerned with providing user discovery for anonymity networks (see \S\ref{sec:related-work}). 
Two prominent private user discovery protocols are Alpenhorn~\cite{lazar2016alpenhorn} and UDM~\cite{chaum2021udm}. 
Neither of these protocols can hide whether a given username is a member of the network, and they cannot maintain availability in the presence of even a single faulty server. 
Moreover, Alpenhorn requires users to regularly download inboxes, whether or not they are performing a lookup.
Consequently, this protocol can lead to substantial monthly bandwidth usage, posing a particular challenge for mobile devices relying on cellular networks. 

In Tor, users can access onion services, previously known as hidden services \cite{onionservices, tor}, through special URLs.
A Tor onion service URL is a 56-character pseudorandom string derived from a public key \cite{onionservices}, making it difficult to remember or communicate verbally.
One can create a shortcut for this URL, for example by using a DNS alias or a service that redirects from a short and memorable URL to the hidden service URL. 
However, this requires the service providing that redirection to be trusted to respond correctly.

In this paper, we propose a practical and metadata-private user discovery protocol named Pudding\footnote{A wordplay on the abbreviation of `private user discovery' (PUD)}, which has two main privacy properties. 
Firstly, Pudding hides contact relationships between users, so that neither a global network observer nor a compromised discovery server can determine who is talking to whom. 
We build our system as an application on top of Loopix \cite{loopix}, which is a practical anonymity network design that has been implemented and is currently operating at significant scale, with hundreds of mix nodes provided as part of the Nym\footnote{\url{https://nymtech.net/}} anonymity network.

Pudding also ensures that an adversarial user cannot learn whether a given username is registered on Pudding, except when authorised by the owner of that username. 
We achieve this by establishing a new mechanism to generate single-use reply blocks (SURBs), which are special network packets that allow a user to send a message to another user without knowing her network address or public key \cite{mixminion,danezis2009sphinx}.
This mechanism allows us to distribute the discovery operations among multiple servers (see \S\ref{sec:deterministic-surb-gen}).

Pudding also ensures that even when up to one third of the directory servers are unavailable or return incorrect responses, a user's request to initiate contact with another user resolves correctly and the system remains operational.

Users who want be discovered through Pudding register and validate access to an email address. 
This allows messaging apps to automate the process of resolving user-readable names into the required network information and cryptographic key material to establish end-to-end encrypted communication.
Moreover, it provides users with reassurance that the person they are talking to is the intended recipient, since the email addresses used come from their own contact address book.
Furthermore, institutional email addresses (e.g.\ \texttt{famous.journalist@newspaper.com}) allow the sender of a message to assure themselves that they are talking to someone at a particular organisation. 

Using a non-anonymous identifier such as a phone number or an email address may appear counter-intuitive in an anonymity network.
However, our design does not require users to register in order to be able to discover other users.
Therefore, for instance, a journalist may publish their Pudding username as part of their social media presence, and whistleblowers can use Pudding to discover and contact the journalist securely, without having to register to Pudding with their own email address.

In this paper, we make the following contributions:

\begin{enumerate}
    \item We identify the need for a new \textit{usable} and \textit{private} user discovery mechanism in anonymous messaging systems and define usability and privacy requirements in this context (\S\ref{sec:problem-definition}).
    \item We propose a new private user discovery protocol named \textit{Pudding} (\S\ref{sec:user-discovery}) that conceals who is talking to whom, tolerates Byzantine or unavailable servers, does not require users to be constantly online, protects the information that a certain email address belongs to a member of the network or not, and supports authenticated usernames that can be short and memorable.
    \item We propose a new method for generating single-use reply blocks (SURBs) \cite{mixminion,danezis2009sphinx}, which allows the servers to collectively respond to a discovery query without needing to coordinate.
    By comparing the responses from several nodes we provide resilience against malicious or compromised discovery nodes, as well as concealing registered usernames.
    \item We show that our protocol is practical by evaluating a prototype over the Nym anonymity network. Registration has a mean latency of less than 9\,seconds, and end-to-end discovery takes less than 13\,seconds on average. These numbers are dominated by delays in the underlying anonymity network.
\end{enumerate}

\section{Problem definition}\label{sec:problem-definition}

To establish contact with a user on an anonymity network, one needs to know some network-specific information about that user.
We use the term \textit{contact information} to refer to the information that is needed to make initial contact with a user through the anonymity network.
Although the content of contact information differs among anonymity network schemes, it generally includes a user's public key. 
In some systems, such as Talek \cite{talek} and Vuvuzela \cite{van2015vuvuzela}, the public key is used as a user ID as well. 
In Loopix \cite{loopix}, the contact information is comprised of the public key, the IP address of the user's provider, and the account ID at that provider (see \S\ref{sec:setup}). 
Nym is similar to Loopix, only using a Nym-specific provider ID instead of an IP address. 

We define \textit{user discovery} as initiating contact with a user based on her username, either by obtaining her contact information, or by directly sending a message to the user without revealing her contact information.
We believe that user discovery should be simple, so that it is accessible to all users. 
Inspired by trust establishment usability definitions from Unger et al.'s work \cite{unger2015sok}, we identify two key usability requirements for user discovery.
A \textit{usable} discovery mechanism in an anonymity network should:
\begin{enumerate}
    \item Require only a short, memorable, and hence low-entropy username to establish contact. The username can either be a `pseudonym' that is unique only within the system, or an identifier that has meaning outside of the anonymity network and therefore may already be in the local address books of the user's friends (e.g.\ a phone number or an email address).
    \item Allow the user discovery to happen without requiring the user to leave the anonymity network, because any communication outside the anonymity network risks revealing to a network adversary the very metadata that the anonymity network is trying to hide.
\end{enumerate}

A private user discovery mechanism must not undermine the level of privacy provided by the anonymity network.
Therefore, a user discovery system should ensure that:
\begin{enumerate}
    \item The information ``Alice is searching for Bob'' is kept secret from everyone other than Alice and Bob.
    The providers of the discovery service and anonymity network, and any active or passive network adversaries, should not learn which users are communicating.
    \item A user should not be able to learn whether a given username is registered, except by contacting the owner of that username and the recipient choosing to respond.
\end{enumerate}

As we show in \S\ref{sec:related-work}, there is no existing solution for user discovery that fits the usability and privacy needs of anonymity networks, is a lightweight design that is compatible with applications running on mobile devices, and provides fault tolerance.
\section{System overview and goals}\label{sec:system-goals}

\begin{figure}[htb]
    \centering
    \includegraphics[width=0.5\textwidth]{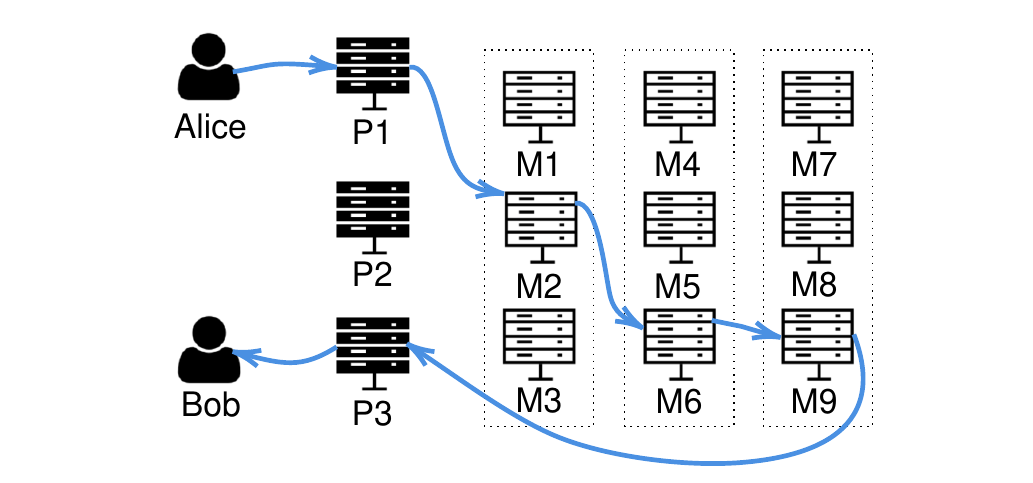}
    \caption{Diagram showing Alice sending Bob a message through a Loopix network with three layers of mix nodes \textit{(M1--M9)} and three provider nodes \textit{(P1, P2, P3)}. The arrows show a possible route for Alice's message. Each message passes through one mix node from each layer. Alice and Bob access the network via provider nodes, which are connected to all the mix nodes in the outermost layers.}
    \label{fig:loopix}
\end{figure}

\subsection{Setup}\label{sec:setup}

\point{Loopix network architecture}
Pudding builds upon the Loopix~\cite{loopix} design.
Loopix is a prominent medium-latency mix network \cite{chaum1981untraceable} that protects sender-receiver relationships from third parties, including global passive and active adversaries, enabling metadata-private email or instant messaging applications.
Loopix hides communication patterns using cover traffic and Poisson mixing.
The network consists of three types of nodes, as illustrated in Figure \ref{fig:loopix}:
(1)~\emph{user devices} are associated with a given user and provide send/receive functionality for end-user applications;
(2) multiple layers of \textit{mix nodes} transmit real and cover messages;
(3) \emph{provider nodes} manage access to the network.
Each user device is registered at a provider node and can only send and receive packets through its provider.
A provider node can receive messages on the user's behalf, even when the user device is offline.
Loopix network packets use the Sphinx packet format, which we describe below.

Pudding is an application-layer protocol~-- it does not require any modification of the anonymity network for which it provides a user discovery service. 
Pudding can be built upon any anonymity network that allows messaging without explicitly knowing the recipient's contact information (see \textit{single-use reply block} below), such as Nym. 
In this paper, we build Pudding on Nym/Loopix, because this anonymity network design provides a high level of metadata privacy, is suitable for asynchronous messaging, and is designed to be feasible to use on mobile devices. 

\point{Pudding nodes}For our protocol, we introduce a new type of server, the \emph{discovery node}, which provides the user discovery service.
Discovery nodes can be independently managed and maintained, and no single entity should control more than one-third of the discovery nodes (see \S\ref{sec:threat_model}).
There is a fixed set of $n$ discovery nodes, where $n=4$ or $n=7$ would be reasonable choices (see \textit{Security parameters} in \S\ref{sec:user-discovery}).
Each discovery node has a database that maps usernames to contact information.
From a Loopix protocol point of view, discovery nodes act as user devices (although in practice, they may be co-located with mix nodes or provider nodes).
This allows discovery nodes to be added to the network without changes to the Loopix protocol, without disturbing its security properties.

\point{Sphinx packet format}In Pudding, user devices and discovery nodes communicate via Loopix (i.e.\ by routing each message through all mix layers) using the Sphinx \cite{danezis2009sphinx} packet format. 
The one exception to this rule is the validation of email addresses, as discussed in \S\ref{sec:registration}.
A Sphinx packet consists of two main components: a header and an encrypted payload. 
The header contains the metadata necessary to route and process the message correctly, i.e.\ to verify the integrity of the packet at each hop, to send the packet to the next node in the path, and to decrypt the outer-most payload layer.
The layered encryption ensures that each node in the message path learns only the routing information to the next hop and a network observer cannot correlate a mix node's incoming and outgoing packets.
Sphinx packets are source-routed, meaning that the creator of the header decides and encodes the routing information into the packet \cite{danezis2009sphinx}.

\point{Single-use reply block (SURB)} Pudding uses a special Sphinx header called a \textit{single-use reply block (SURB)} for some of its messages. 
A SURB is a Sphinx header that allows a user to send a message without learning the identity of the recipient~\cite{mixminion,danezis2009sphinx}.
SURBs are typically used when user Alice wants to send a message to Bob and allow him to reply to her without revealing her own contact information.
Alice constructs a Sphinx header with onion-encrypted routing information that goes through a sequence of mix nodes and ends at herself.
She can then include this SURB. which includes keys for the payload encryption, in her message to Bob.
While SURBs are often used for anonymous replies to the original sender, they can be constructed by anyone with knowledge of the contact information of the destination.

\point{System assumptions}
We assume that all user devices know the public contact information for all discovery nodes, so that they can send anonymous messages to the discovery nodes.
Provider nodes can supply this information to user devices.

\subsection{Threat Model and Goals}\label{sec:threat_model}

\point{Security and privacy goals}\label{sec:sec-priv-goals}
Our security and privacy goal definitions are inspired by AnoA~\cite{backes2013anoa}, which is also the approach taken by Loopix~\cite{loopix}. 
In terms of notation, each user ($U$) possesses a username ($\mathit{ID}$) and contact information ($\Delta$). We denote the $\mathit{ID}$ and $\Delta$ associated with a certain user $U$ as $\mathit{(U:ID,\,} \Delta \mathit{)}$.
We write $\{U_i \rightarrow U_j\}$ to indicate that user $U_i$ searches for user $U_j$.
For now we state our goals informally, and we formalise them in \S\ref{sec:security-analysis}.

\begin{enumerate}[label={\bfseries G\arabic*}]
    \item \textbf{Unlinkability:}\label{goal:unlinkability}
    The adversary cannot learn \textit{who is searching for whom}. 
    This goal ensures that the \textit{sender-receiver third-party unlinkability} goal of Loopix is preserved.
    In other words, assuming a sufficiently large Loopix anonymity set, the adversary cannot tell the difference between $\{U_1 \rightarrow U_*\}$ and $\{U_2 \rightarrow U_*\}$, for all honest users $U_1$ and $U_2$ chosen by the adversary.
    Essentially, when a user searches for $U_*$, the adversary learns $U_*$ but not who sent the query.
    \item \textbf{Security against impersonation:}\label{goal:impersonation}
    Assume a user $U_1$ has registered contact information $\Delta_1$ for a particular username $\mathit{ID}_1$. 
    When another user subsequently looks up $\mathit{ID}_1$, the contact request should be sent to the registered contact information $\Delta_1$ and no other user. 
    In simpler terms, an adversary cannot cause any honest user to send a message intended for $U_1$ to a $\Delta_2 \neq \Delta_1$.
    \item \textbf{External identity verification:}\label{goal:id_verification}
    A user's identity within the system should be verifiably associated with an external identity by the discovery nodes.
    For instance, this can be achieved by discovery nodes sending a code to an email address or phone number, assuming that the communication channel used to send the code is secure and the user's device is trusted. 
    This goal provides assurance to the user that she is talking to the correct person, defined as the owner of an email address.
    Let us assume that the user $U$ is a user controlled by the adversary, and $\mathit{ID}$ is an external identity, such as an email address at a trusted email service, that is \emph{not} controlled by the adversary.
    We define external identity verification as the inability of the adversary to convince any honest discovery node into registering the user $(\mathit{U:ID},\, \Delta)$.
    \item \textbf{Membership unobservability:}\label{goal:membership_unobservability}
    No user should be able to determine whether a username has been registered, unless the owner of that username decides to reveal this information. 
    Membership unobservability is important in situations where exposing the fact that someone is a member of a network may harm the user.
    For a username $\mathit{ID}$ chosen by the adversary, we define \textit{membership unobservability} as the inability of the adversary to distinguish with better than negligible probability whether there exists some $\mathit{U},\, \Delta$ such that $(\mathit{U:ID}, \Delta)$ is a registered honest user.
\end{enumerate}

\point{Threat Model}
For the \textit{unlinkability} (\textbf{\ref{goal:unlinkability}}), \textit{security against impersonation} (\textbf{\ref{goal:impersonation}}), and \textit{external identity verification} (\textbf{\ref{goal:id_verification}}) goals, we assume the following threat model.
We assume that the adversary can join Pudding with any number of user devices, and may observe the network traffic globally, or interfere with it in arbitrary ways.
We assume that at most $f$ out of $n = 3f+1$ discovery nodes are faulty.
A faulty node is either non-Byzantine (crashed, disconnected), or Byzantine (may maliciously deviate from the protocol).
We assume that Byzantine nodes are controlled by the adversary, and that they may collude.
Non-Byzantine nodes follow an honest-but-curious model~\cite{Paverd2014}: they correctly follow the protocol, but any messages they receive may be leaked to the adversary.
The set of non-Byzantine-faulty nodes may change over time, whereas a Byzantine node is assumed to remain Byzantine forever.
The users and the non-faulty nodes do not need to know which nodes are faulty.
This is a common approach for Byzantine fault-tolerant protocols~\cite{Lamport:1982fr}.

We assume that the underlying Loopix construction protects against active and passive network adversaries who wish to learn which nodes are communicating with each other.
We base these assumptions upon the Loopix threat model~\cite{loopix}. 
Lastly, we assume that (1) the email services employed for username verification ensure that an email originating from a specific email address contains a DKIM signature (see \S\ref{sec:id-verified-core-idea}) corresponding to that email address, and (2) that email traffic is encrypted with TLS so that a network adversary cannot read or modify messages.

For the \textit{membership unobservability} (\textbf{\ref{goal:membership_unobservability}}) goal, we assume a weaker threat model, where the adversary controls any number of user devices but no mix, provider, or discovery nodes (an adversary with access to a single discovery node's database can trivially determine which usernames are registered).
Despite this weaker threat model, providing membership unobservability against user devices significantly limits the number of parties that have access to this information, and also prevents crawling the database of usernames, which has been an issue in existing encrypted messaging apps \cite{hagen2021all}. 

\point{Security parameters}
We assume $n=3f+1$ discovery nodes for the following reason.
Assume $f$ discovery nodes are temporarily unavailable during registration, so $n-f$ discovery nodes have a copy of the contact information for a given username.
When searching for that username, a different set of $f$ discovery nodes may be unavailable, but at least $n-2f\ge f+1$ nodes will be both available and have a copy of the contact information of that username, which will be sufficient to allow discovery of this username.

\section{User discovery}\label{sec:user-discovery}

Pudding user discovery allows a user device to make contact with the registered owner of a particular username.
Registering to Pudding with a username allows a user to become discoverable through Pudding by that username, but a user who wants to search for another user does not need to register with Pudding.
We first discuss the discovery protocol, and defer the description of the registration protocol to \S\ref{sec:registration}.

\subsection{Deterministic SURB generation}\label{sec:deterministic-surb-gen}
Pudding's \textit{membership unobservability} goal (see \S\ref{sec:system-goals}) requires that a searcher cannot tell from the discovery nodes' responses whether a given username has been registered.
We achieve this by always returning a SURB for a discovery request. 
If the searched username is registered at the discovery nodes, the SURB will allow the searcher to send a message to the username's owner.
Otherwise, the result of the discovery query is a fake SURB that does not route to any real user, but simply causes the message to be dropped in the mix network.
This provides membership unobservability because (i)~a user cannot distinguish whether a SURB was generated with fake or real contact information, and (ii)~a user cannot distinguish whether a message was dropped by the network, or the recipient user chose not to respond to the message.
Although the final mix node in a packet's route can tell whether a SURB is fake, as per \S\ref{sec:threat_model}, the threat model for membership unobservability only includes malicious users, not adversarial mix or discovery nodes.

SURB generation cannot simply be performed by a single discovery node, since otherwise that discovery node could impersonate any user by routing the searcher's message to an adversary-controlled destination.
Instead, we define a deterministic scheme for generating SURBs, so that all honest discovery nodes generate the same SURB for the same query.
We then require that a user receives an identical SURB from at least $f+1$ distinct discovery nodes before sending a message using that SURB.
Since we assume that a maximum of $f$ discovery nodes may be Byzantine, there is no way that the adversary can cause the user to receive $f+1$ incorrect SURBs, and therefore any SURB received from at least $f+1$ discovery nodes must be correct.

\point{Traditional SURB generation} 
The original SURB generation procedure for Sphinx \cite{danezis2009sphinx} takes two inputs: (1) a destination contact information $\Delta$ and (2) a sequence of mix nodes $\{n_0, n_1, \dotsc, n_{v-1}\}$ through which the message will be routed, where $v < r$ and $r$ is the maximum number of mix nodes in a message route.
The SURB generation procedure in Loopix needs randomness for two tasks: 
(1) picking a delay value $d_i$ for the message to be delayed at each hop $n_i$, chosen from an exponential distribution so that each mix node behaves as a Poisson process; and (2) picking an element of a prime-order group, which is used by each mix node in the path to derive a shared secret key, known by the mix node and the SURB's creator.
Each mix node uses this shared secret key to derive the keys for validating the integrity of the message, removing a layer of encryption, and learning the next hop in the message route \cite{danezis2009sphinx, loopix}. 

\point{Deterministic SURB generation}
Our SURB generation process is identical to the original procedure, with the exception that all sources of randomness are replaced with a pseudorandom generator seeded with $\mathsf{KDF}(\mathit{nonce} \parallel \mathit{username} \parallel k)$, where $\mathit{username}$ is the username being queried, $\mathit{nonce}$ is a nonce in the query message, $k$ is a long-lived secret that is shared among all the discovery nodes (set up when the discovery nodes are created), and $\mathsf{KDF}$ is a key derivation function.
Instead of taking the sequence of mix nodes $\{n_0, n_1, \dotsc, n_{v-1}\}$ (the message route) as an input, it also generates this sequence pseudorandomly from the KDF output.
The inputs to our modified procedure are therefore the $\mathit{username}$, the $\mathit{nonce}$, and the destination contact information $\Delta$.
Determinism ensures that all honest discovery nodes generate the same SURB for the same destination contact information and nonce.
Since $k$ is not known to users, a malicious user cannot run the same SURB generation algorithm to break membership unobservability.

While not explicitly outlined in our protocol, potential leaks of the $k$ value can be mitigated through an epoch-based mechanism, where the discovery nodes periodically agree on fresh $k$ values. 
Lookup failures during key rotation can be reduced by having overlapping epochs.
The epoch number can also be incorporated into nonces, which avoids having to store nonces forever.

\point{Creating fake SURBs}
When a user queries a username that is not registered on the system, Pudding nevertheless returns a valid SURB in order to provide membership unobservability.
This SURB is generated in the same way as a SURB for a genuine user, except that the discovery nodes use a fixed false contact information $\Delta_\mathsf{fake}$ instead of $\Delta$.
This $\Delta_\mathsf{fake}$ is a non-existent ``black hole'' destination, so that any messages sent to it are dropped by the mix network.
The querying user cannot distinguish a SURB for the black hole from a SURB for real contact information, because the outermost layer of the SURB header is encrypted with the public key of the first mix node on the path \cite{danezis2009sphinx}, and thus cannot be decrypted by the user.

If a user makes queries for two different usernames that are both unregistered, the SURB responses will be constructed based on the same $\Delta_\mathsf{fake}$ destination.
However, since the seed of the pseudorandom generator incorporates the queried username, the SURBs are encrypted with different keys, and therefore the querying user cannot tell whether the SURBs are based on the same or different contact information.
Having a single, fixed $\Delta_\mathsf{fake}$ eliminates the need for coordination between discovery nodes.

The public key within $\Delta_\mathsf{fake}$ must be chosen in a way that the corresponding private key is not known by any party in the system.
For elliptic curves, this can be achieved by running a hash-to-curve algorithm \cite{hashtocurve} with a simple constant input, such as the string ``Pudding fake identity'', to obtain a group element whose discrete logarithm is unknown.

\subsection{Authenticating contact information}\label{sec:blinded_key_signature}
The SURB described in the previous section allows Alice to send a single message to Bob without Alice learning Bob's contact information, and without Bob learning Alice's identity.
However, if Alice wishes to identify herself to Bob by including her username in her message, the SURB does not allow Bob to verify that the message did indeed come from Alice, nor does it allow Alice to verify that replies to her message did indeed come from Bob.
Security against impersonation (\ref{goal:impersonation}) requires users to authenticate each other.

If membership unobservability were not a concern, the discovery nodes could return a user's public key (which is part of their contact information $\Delta$) when queried for a username.
Alice and Bob could query the discovery servers for each other's usernames to learn each other's public keys, and then perform an authenticated key exchange protocol to establish a mutually authenticated channel.

To enable mutual authentication while preserving membership unobservability, we use key-blinded signatures \cite{keyblindingsignatures, eaton2021post}, a technique also used by Tor to maintain anonymity of Tor onion services \cite{onionservices, hopperproving}.
(Not to be confused with blind signatures \cite{chaum1983blind}: key-blinded signatures hide the unblinded public key from the verifier, whereas blind signatures hide the message from the signer.)

The scheme we use \cite{keyblindingsignatures} works as follows.
Let $g$ be a generator of an elliptic curve group with prime order $p$.
Let $x \in \mathbb{Z}_p$ be a private key and let $g^x$ be the corresponding public key. 
Let $y \in \mathbb{Z}_p$ be a blinding factor chosen uniformly at random.
One can then sign and verify digital signatures similarly to regular ECDSA or EdDSA signatures, using $xy \pmod{p}$ as private key and $g^{xy}$ as public key. %
Given a user's public key $g^x$, a discovery node can therefore generate blinded public keys $(g^x)^{y_1}$, $(g^x)^{y_2}, \dots$ that cannot be linked with each other, or with the unblinded public key $g^x$, without knowledge of the blinding factors $y_1, y_2, \dots$ \cite{eaton2021post}.
We write the blinded-key signing operation as $\mathsf{SignBK}(\mathit{sk}, \mathit{bk}, \mathit{msg})$ where $\mathit{sk}$ is the unblinded private key, $\mathit{bk}$ is the blinding key, and $\mathit{msg}$ is the data to be signed.

\renewcommand{\floatpagefraction}{.8}%
\begin{figure}
  \begin{subfigure}{\linewidth}
    \includegraphics[width=\linewidth]{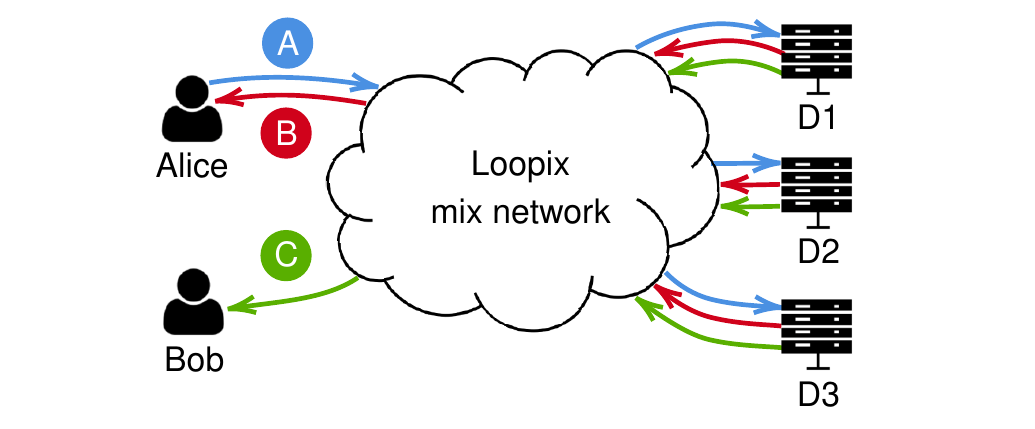}
    \caption{During the \lookup protocol: (A) Alice sends a query message containing Bob's username to each discovery node; (B) each discovery node replies with a SURB and a blinded public key for Bob; (C) each discovery node also sends Bob the blinding key.}
  \end{subfigure}%
  \\
  \begin{subfigure}{\linewidth}
    \includegraphics[width=\linewidth]{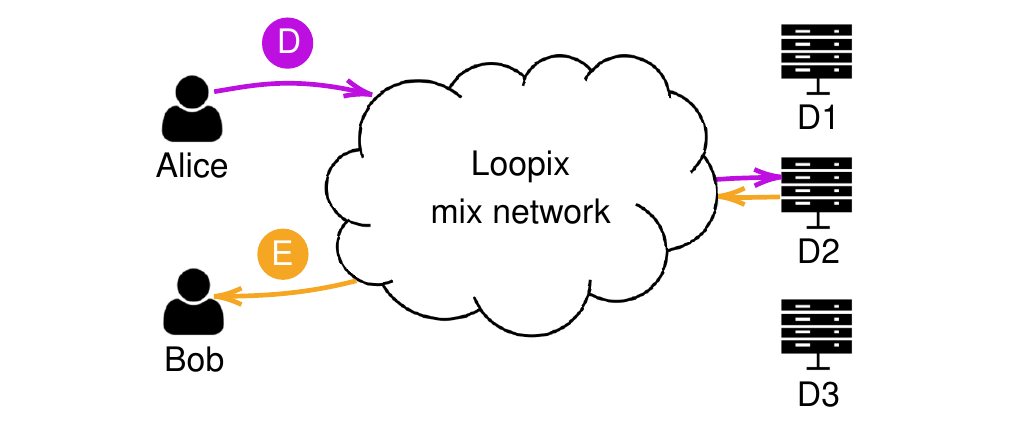}
    \caption{Then, Alice runs the \continit protocol where: (D) Alice creates a Sphinx packet $M_\mathsf{init}$ using the SURB she received in \lookup. Alice then sends a message to an arbitrary discovery node, where $M_\mathsf{init}$ is the payload. (D) The discovery node places $M_\mathsf{init}$ in the network, which routes the message to Bob.}
  \end{subfigure}%
\caption{Alice discovering Bob (\lookup) and and initiating contact with him (\continit) through a Pudding setup with three discovery nodes \textit{(D1, D2, D3)}.} \label{fig:discovery}
\end{figure}

\subsection{Discovery}
\label{sec:lookup}
User discovery in Pudding consists of three protocol phases: \lookup queries the database of users, \continit allows the searcher to send an initial message to the queried user, and \addfriend establishes an authenticated channel between the two users.

\point{\lookup protocol}
Let Alice be a user who is searching for Bob, and let $\mathit{ID}_\mathsf{A}$ and $\mathit{ID}_\mathsf{B}$ be their usernames (having a username is optional in Alice's case).
Note that we use ``Alice/Bob'' as shorthand for ``Alice's/Bob's user device''.
Alice and Bob are both Loopix users, with contact information $\Delta_\mathsf{A}$ and $\Delta_\mathsf{B}$ respectively.
Each user's contact information includes a long-term public key for which the user's device holds the corresponding private key.
Alice's private key is $x_\mathsf{A} \in \mathbb{Z}_p$ and her public key is $g^{x_\mathsf{A}}$, whereas Bob's private and public keys are $x_\mathsf{B}$ and $g^{x_\mathsf{B}}$ respectively.
We also assume that each discovery node has a private signing key, and that Alice and Bob know the corresponding public key for each discovery node (this information can be part of the network topology that a user downloads when first connecting to the Loopix network).
The protocol is illustrated in Figure~\ref{fig:discovery}:

\begin{enumerate}
     \item \label{item:select-discovery-nodes}Alice generates a random $\mathit{nonce}$.
     For each discovery node $i$, Alice also generates a SURB $S_i$ that allows that discovery node to reply to her.
     She then sends a discovery message $(\mathit{ID}_\mathsf{B}, \mathit{nonce}, S_i)$ to each discovery node via the anonymity network.
    \item \label{item:discovery-surb} Upon receiving a discovery message, each discovery node checks if $\mathit{nonce}$ has been used before; if so, it drops the message.
    This prevents a user from detecting when the contact data for another user changes by making repeated queries using the same nonce.
    \item\label{item:user-discovery-surb-reply} Each discovery node seeds a pseudorandom generator with $\mathsf{KDF}(\mathit{nonce} \parallel \mathit{ID}_\mathsf{B} \parallel k)$, where $k$ is a secret shared with the other discovery nodes as described in \S\ref{sec:deterministic-surb-gen}.
    If the discovery node has a record for $\mathit{ID}_\mathsf{B}$ with contact information $\Delta_\mathsf{B}$ in its database, it creates a deterministic SURB $\mathit{surb}$ using $\Delta_\mathsf{B}$ and the pseudorandom generator; if not, it generates the SURB using $\Delta_\mathsf{fake}$ instead.
    Each discovery node also obtains a blinding key for Bob $y_\mathsf{B} \in \mathbb{Z}_p$ from the pseudorandom generator, obtains Bob's public key $g^{x_\mathsf{B}}$ from $\Delta_\mathsf{B}$, and computes the blinded public key $\mathit{bpk}_\mathsf{B} = (g^{x_\mathsf{B}})^{y_\mathsf{B}}$.
    If $\mathit{ID}_\mathsf{B}$ is not registered, it computes a blinded key $\mathit{bpk}_\mathsf{B} = (\mathit{pk}_\mathsf{fake})^{y_\mathsf{B}}$ based on the public key $\mathit{pk}_\mathsf{fake}$ in $\Delta_\mathsf{fake}$.
    \item Each discovery node computes a signature $\mathit{sig}_\mathsf{A} = \mathsf{Sign}_{sk}(\mathit{nonce} \parallel \mathit{surb} \parallel \mathit{bpk}_\mathsf{B})$ using its signing key $sk$, and sends $(\mathit{nonce}, \mathit{surb}, \mathit{bpk}_\mathsf{B}, \mathit{sig}_\mathsf{A})$ to Alice using the SURB $S_i$ in Alice's discovery message.
    \item If $\mathit{ID}_\mathsf{B}$ is registered with contact information $\Delta_\mathsf{B}$, each discovery node also computes a signature $\mathit{sig}_\mathsf{B} = \mathsf{Sign}_{sk}(\mathit{nonce} \parallel y_\mathsf{B})$ using its signing key $sk$, and sends $(\mathit{nonce}, y_\mathsf{B}, \mathit{sig}_\mathsf{B})$ to Bob using $\Delta_\mathsf{B}$.
    This will allow Bob to generate signatures for the blinded public key.
    \item \label{item:lookup-receive-surbs}
    For each message received, Alice checks that $\mathit{sig}_\mathsf{A}$ is a valid signature by one of the discovery nodes.
    She waits until she has received the same $\mathit{surb}$ and $\mathit{bpk}_\mathsf{B}$, along with the $\mathit{nonce}$ she generated, from at least $f+1$ distinct discovery nodes.
    If she does not receive them within some timeout, she discards all values and restarts the protocol.
    \item \label{item:lookup-receive-blinding-key} 
    For each message received, Bob checks that $\mathit{sig}_\mathsf{B}$ is a valid signature by one of the discovery nodes.
    If Bob receives $f+1$ copies of the same blinding key $y_\mathsf{B}$ and $\mathit{nonce}$ from distinct discovery nodes, he stores the two values locally in anticipation of a future contact request (see \continit Step \ref{item:continit-reply-ignore}).
\end{enumerate}

\point{\continit protocol}
The SURB received from $f + 1$ discovery nodes during the \lookup protocol allows Alice to send a message to Bob anonymously.
However, if Alice sends her initial message directly to Bob using this SURB, an adversary who controls a discovery node in addition to Alice's provider or the first mix node on the message path can recognise the SURB and link it to Alice. 
To prevent such linking attacks, the message is reflected via a different node.
For simplicity, we choose a discovery node as the reflector.
Alternatively, the system can incorporate separate, untrusted reflector nodes, or even allow other users to serve as message reflectors.
Our protocol allows Alice to incorporate a pseudonym or a codeword in her initial message, enabling Bob to make an informed decision about whether to respond to her.

\begin{enumerate}
    \item \label{item:continit-keygen} Alice generates an ephemeral Diffie-Hellman private key $a \in \mathbb{Z}_p$, public key $g^a$, and initial message key $K_e = \mathsf{KDF}((\mathit{bpk}_\mathsf{B})^a \parallel \text{``init key''})$ from the blinded public key received in \lookup Step \ref{item:lookup-receive-surbs}.
    She then uses the SURB from \lookup Step \ref{item:lookup-receive-surbs} to create a Sphinx packet $M_\mathsf{init}$ \cite{danezis2009sphinx}.
    The payload of $M_\mathsf{init}$ contains $g^a$, the nonce value she used when searching for Bob, as well as the following fields encrypted under $K_e$: (1) a SURB that allows Bob to respond to her, (2) an optional pseudonym or codeword for Bob, and (3.a) either her username $\mathit{ID}_\mathsf{A}$, or (3.b) if she is not registered or does not want to share her username, Alice samples a blinding key $y_\mathsf{A} \in \mathbb{Z}_p$ and uses her long-term public key $g^{x_\mathsf{A}}$ to compute a blinded public key $\mathit{bpk}_\mathsf{A} = (g^{x_\mathsf{A}})^{y_\mathsf{A}}$, which she includes in $M_\mathsf{init}$.
    Alice then picks a discovery node and sends it a message via the anonymity network that contains $M_\mathsf{init}$ as a payload.
    \item The discovery node that receives Alice's message sends $M_\mathsf{init}$ through the anonymity network.
    It is routed to Bob using the SURB created by the discovery nodes.
    \item \label{item:continit-reply-ignore} When Bob receives $M_\mathsf{init}$, he computes $K_e = \mathsf{KDF}((g^a)^{x_\mathsf{B} y_\mathsf{B}} \parallel \text{``init key''})$ using his private key $x_\mathsf{B}$ and the blinding key $y_\mathsf{B}$ received in \lookup Step \ref{item:lookup-receive-blinding-key}, and then decrypts the rest of the message.
    Based on Alice's codeword he can either decide to reply to it, or ignore it. 
    If Bob decides to reply, he proceeds to the \addfriend protocol specified below.
    If Bob decides to ignore the message, Alice does not learn if $\mathit{ID}_\mathsf{B}$ is registered to Pudding or not.
    \item If Alice does not receive a reply from Bob within some timeout, she retries by sending $M_\mathsf{init}$ through a different discovery node up to $f+1$ times to route around faulty discovery nodes. %
\end{enumerate}

\point{\addfriend protocol}
If Bob decides to accept Alice's contact request, the users need to establish an authenticated communication channel: Alice needs to ensure that she is indeed talking to the user registered under the username $\mathit{ID}_\mathsf{B}$, and if Alice supplied her username, Bob needs to check that he is indeed talking to $\mathit{ID}_\mathsf{A}$.

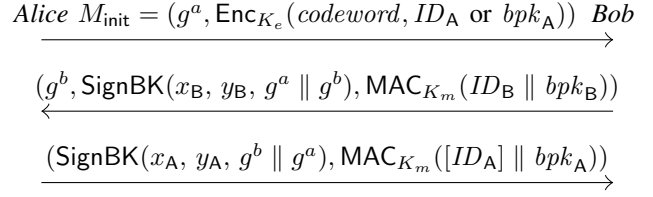
\begin{figure}
    \centering
    \begin{tikzpicture}[scale=0.95]
    \tikzstyle{arrow}=[-{Classical TikZ Rightarrow[length=3pt]}]
    \tikzstyle{label}=[above,text depth=3pt]
    \draw [arrow] (0,2) node [label] {\emph{Alice}} -- node [label] {$M_\mathsf{init} = (g^a, \mathsf{Enc}_{K_e}(\mathit{codeword}, \mathit{ID}_\mathsf{A} \text{ or } \mathit{bpk}_\mathsf{A}))$} (8,2) node [label] {\emph{Bob}};
    \draw [arrow] (8,1) -- node [label] {$(g^b, \mathsf{SignBK}(x_\mathsf{B},\, y_\mathsf{B},\, g^a \parallel g^b), \mathsf{MAC}_{K_m}(\mathit{ID}_\mathsf{B} \parallel \mathit{bpk}_\mathsf{B}))$} (0,1);
    \draw [arrow] (0,0) -- node [label] {$(\mathsf{SignBK}(x_\mathsf{A},\, y_\mathsf{A},\, g^b \parallel g^a), \mathsf{MAC}_{K_m}([\mathit{ID}_\mathsf{A}] \parallel \mathit{bpk}_\mathsf{A}))$} (8,0);
    \end{tikzpicture}
    \caption{Summary of the \addfriend protocol.
    Square brackets denote that including Alice's username is optional.
    Omitted from the diagram: each message also includes the nonce from Alice's original \lookup request (to identify messages belonging to the same protocol run) and a SURB that allows the other party to send a reply message; Bob's message to Alice also includes the nonce he used in his optional \lookup run for Alice's username.}
    \label{fig:addfriend}
\end{figure}

We achieve mutual authentication using the SIGMA authenticated key exchange protocol \cite{krawczyk2003sigma}, adapted to support blinded-key signatures (see \S\ref{sec:blinded_key_signature}).
We use SIGMA because it is fairly simple and proven correct~\cite{canetti2002sigma}, but another authenticated key exchange protocol could be substituted in its place.
The \addfriend protocol is shown in Figure~\ref{fig:addfriend}.

\begin{enumerate}
    \item \label{item:addfriend_lookup}
    Bob obtains a blinded public key $\mathit{bpk}_\mathsf{A}$ for Alice:
    (a) if $M_\mathsf{init}$ does not include Alice's username, Bob uses the $\mathit{bpk}_\mathsf{A}$ in this message;
    (b) if $M_\mathsf{init}$ includes $\mathit{ID}_{\mathsf{A}}$, Bob obtains a blinded public key $\mathit{bpk}_\mathsf{A}=(g^{x_\mathsf{A}})^{y_\mathsf{A}}$ for Alice by running \lookup with input $\mathit{ID}_{\mathsf{A}}$ and using the value returned by at least $f+1$ discovery nodes.
    \item \label{item:addfriend_alice_public_key}
    Bob uses the blinding key $y_\mathsf{B}$ he received from the discovery servers (\lookup Step \ref{item:lookup-receive-blinding-key}) and his public key $g^{x_\mathsf{B}}$ to obtain his own blinded public key $\mathit{bpk}_\mathsf{B} = (g^{x_\mathsf{B}})^{y_\mathsf{B}}$.
    \item \label{item:addfriend_bob_reply} Bob generates an ephemeral Diffie-Hellman private key $b \in \mathbb{Z}_p$ and public key $g^b$, and calculates the MAC key $K_m = \mathsf{KDF}((g^a)^b \parallel \text{``MAC key''})$ using the value $g^a$ from $M_\mathsf{init}$.
    He uses $K_m$ to calculate the message authentication code $\mathsf{MAC}_{K_m}(\mathit{ID}_\mathsf{B} \parallel \mathit{bpk}_\mathsf{B})$.
    Bob also generates $\mathsf{SignBK}(x_\mathsf{B},\, y_\mathsf{B},\, g^a \parallel g^b)$ using the key-blinded signature scheme \cite{keyblindingsignatures}.
    Finally, he sends a message to Alice using the SURB in $M_\mathsf{init}$, containing $g^b$, the MAC, the signature, and a SURB for Alice to reply. 
    If he ran \lookup for $\mathit{ID}_\mathsf{A}$ in Step \ref{item:addfriend_lookup}, Bob also includes the nonce he used in this \lookup.
    \item \label{item:addfriend-alice-blinding} If Alice included her username $\mathit{ID}_{\mathsf{A}}$ in $M_\mathsf{init}$, she waits until she has received a blinding key $y_\mathsf{A}$ from at least $f+1$ distinct discovery nodes, as a result of Bob's \lookup run at Step \ref{item:addfriend_lookup} of \addfriend.
    She uses the nonce in Bob's message to identify the blinding key.
    If Alice chose to remain anonymous, she uses the blinding key $y_\mathsf{A}$ that she generated at \continit Step \ref{item:continit-keygen}.
    \item \label{item:addfriend-alice-check} When Alice receives Bob's message, she checks that the signature over $g^a \parallel g^b$ is valid using the blinded public key $\mathit{bpk}_\mathsf{B}$ she received from the discovery nodes.
    She then calculates the MAC key $K_m = \mathsf{KDF}((g^b)^a \parallel \text{``MAC key''})$ and uses it to check the MAC over $\mathit{ID}_\mathsf{B}$ (the username she originally searched for) and the blinded public key $\mathit{bpk}_\mathsf{B}$ from the discovery nodes.
    \item \label{item:addfriend_alice_reply} If all checks pass, Alice sends a reply to Bob containing a key-blinded signature $\mathsf{SignBK}(x_\mathsf{A},\, y_\mathsf{A},\, g^b \parallel g^a)$ computed from her private key $x_\mathsf{A}$ and the blinding key $y_\mathsf{A}$ from Step \ref{item:addfriend-alice-blinding}.
    If Alice sent her username in $M_\mathsf{init}$, the message also includes $\mathsf{MAC}_{K_m}(\mathit{ID}_\mathsf{A} \parallel \mathit{bpk}_\mathsf{A})$, where $\mathit{bpk}_\mathsf{A} = g^{x_\mathsf{A} y_\mathsf{A}}$.
    If Alice is anonymous, the message includes $\mathsf{MAC}_{K_m}(\mathit{bpk}_\mathsf{A})$.
    \item When Bob receives Alice's message, he checks the signature using Alice's blinded public key that he obtained in Step~\ref{item:addfriend_lookup}, and checks that the MAC over Alice's blinded public key (and optionally username) is valid using key $K_m$.
    If these checks succeed, Alice and Bob have fully authenticated each other.
    \item Alice and Bob can then derive a shared session key $K_s = \mathsf{KDF}(g^{ab} \parallel \text{``session key''})$, and use an authenticated encryption scheme to secure their further communication (which may include sharing their contact information with each other).
\end{enumerate}

\section{Registration}\label{sec:registration}
A user must register to Pudding with a username to be discoverable through that username. 
However, lookups are possible without registration. 
We describe the Pudding registration protocol in this section.

\subsection{Authentication with DKIM}\label{sec:id-verified-core-idea}
The conventional way of checking whether a user owns a particular email address is by sending a random token to that address and asking the user to send that token back to the server, typically by clicking an HTTPS link in the email. 
However, since Pudding has multiple discovery servers, using traditional email address verification would cause complications. 
The first option to use traditional email authentication in Pudding would be for discovery nodes to trust each other, such that one node performs the verification and the other nodes trust its verdict. 
This option does not meet Pudding's security goals, since we assume that up to $f$ discovery nodes may be Byzantine and hence not perform the email verification truthfully.
The second option would be for each discovery node to perform its own email address verification, meaning the user would have to click $n$ links in $n$ separate emails. 
This is how Alpenhorn handles email verification for its users \cite{lazar2016alpenhorn}.
However, we believe that this method is inconvenient for users. 
We therefore propose a different method for checking whether the user owns an email address.

DomainKeys Identified Email (DKIM) \cite{leiba2007domainkeys} uses a digital signature in a header to confirm the origin of an email.
The recipient can retrieve the public verification key that the domain owner has published as a DNS TXT record and use it to verify the DKIM signature.
The signature confirms that the email was sent via an SMTP server authorised by the owner of the sender's domain \cite{crocker2011domainkeys}.
Assuming that this SMTP server has authenticated the user (the part of the email address before the `@' sign), DKIM allows us to confirm that the sender of a message has not been spoofed.

In Pudding, one discovery node collects challenge values from all discovery nodes, and sends these challenges in a single email to the email address picked as the username.
To confirm receipt, the registering user responds to this email using the reply function of their email client.
Since email applications attach the original message to the response, and servers and automatically sign the email with the domain's DKIM key, the user typically does not have to type or copy-paste anything. %
The discovery node that receives the reply then shares it with all other nodes.
This mechanism allows each discovery node to independently confirm that the user's response email is (1) a genuine registration request (by checking if its challenge value is included in the email) and (2) sent by the owner of the email address (by checking the validity of the DKIM signature). 

\point{Substitutability of DKIM} 
Although this implementation uses DKIM, it is not the only viable verification method. 
For instance, if DKIM verification fails, the system can fall back to a method where each discovery node sends the user a verification link, with the usability downside of requiring multiple confirmation emails and link clicks.

\begin{figure}
    \centering
    \includegraphics[width=\linewidth]{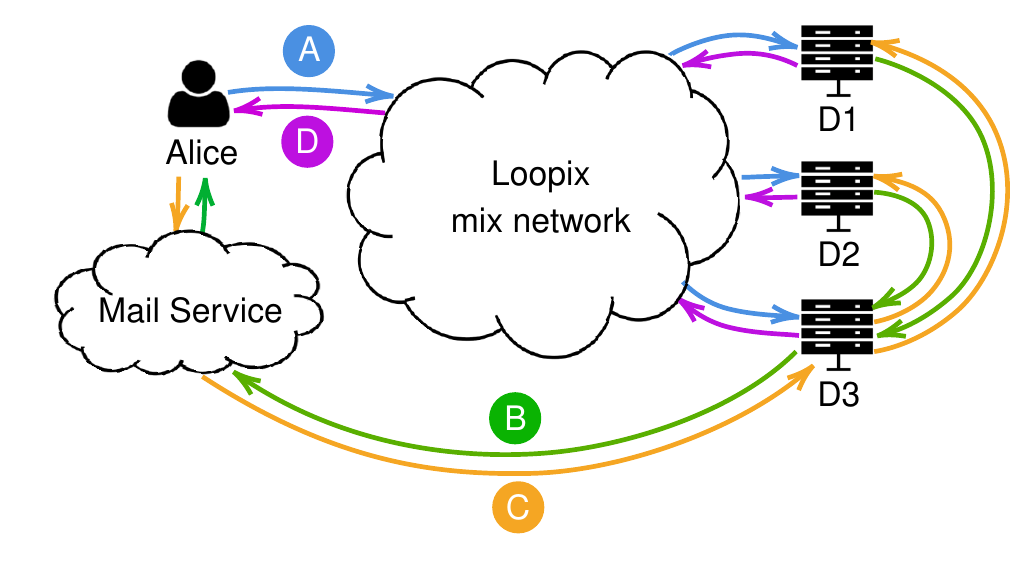}
    \caption{Schematic of user Alice registering to a Pudding setup with three discovery nodes (\textit{D1, D2, D3}). (A) Alice sends to the discovery nodes a message containing her email address, contact information, and the ID of a randomly-selected discovery node $D_\mathsf{auth}$. (B) $D_\mathsf{auth}$, (\textit{D3} in this case), collects challenge values from all discovery nodes and sends Alice an email including these challenges, along with her contact information and username. (C) Alice replies to the email through an email service; \textit{D3} receives Alice's DKIM-signed reply email and sends it to the other discovery nodes. (D) Each discovery node verifies the DKIM signature and its own challenge's presence in the email. Once confirmed, they register Alice and send her a confirmation message.}
    \label{fig:registration}
\end{figure}

\subsection{Registering to Pudding}\label{sec:verified-protocol}

The user registration protocol $\textsc{Register}$ is described below and illustrated in Figure~\ref{fig:registration}.

\begin{enumerate}
    \item The user device joins the Loopix network by signing up to a provider node.
    This process involves creating a public/private key pair for secure communication over the network and obtaining the network topology from the provider (including the set of discovery servers and their public keys). 
    The public key, along with the provider details and any other necessary information, constitutes the user's contact information $\Delta$.
    The user also chooses an email address that she controls as her username $\mathit{ID}$.

    \item \label{item:user-pick-d-auth}The user device randomly picks a discovery node $D_\mathsf{auth}$ as the node responsible for sending the verification email.
    The user then sends a \textit{registration message} to all discovery nodes, which contains $\mathit{ID}$, the contact information $\Delta$, and the identifier of $D_\mathsf{auth}$.

    \item \label{item:first-node-check-id}After receiving a registration message, each discovery node checks whether it has $\mathit{ID}$ already registered, in which case it marks the registration request as invalid. 
    Regardless, the discovery node generates a random challenge value.
    
    \item Any discovery node that is not $D_\mathsf{auth}$ sends its challenge value and $\mathit{ID}$ to $D_\mathsf{auth}$. 
    Meanwhile, $D_\mathsf{auth}$ waits until it has challenges from at least $2f+1$ distinct discovery nodes, including itself (see \S\ref{sec:threat_model}).
    $D_\mathsf{auth}$ then sends a \textit{verification email} to the email address provided as $\mathit{ID}$, which contains all the challenges and Alice's contact information $\Delta$.
    Including $\Delta$ in the email prevents a malicious $D_\mathsf{auth}$ from providing a $\Delta_\mathsf{bad} \neq \Delta$ to the remaining discovery nodes.
    
    \item  If the user receives the email within a predetermined timeout period, she verifies that the included contact information $\Delta$ is correct, and responds with an email that includes all challenge values\footnote{This process can be automated in practice, for example by including a link in the email, which redirects the user from the email app to the anonymous messaging app. After automatically checking that the $\Delta$ is correct, the anonymous messaging app can use the OS features to automatically generate a reply email that contains all the email content and headers, and allow user to reply with the click of a button.}. 
    Otherwise, the user returns to Step~\ref{item:user-pick-d-auth} to try registering again with a new $D_\mathsf{auth}$.

    \item \label{item:first-node-dkim-check}When $D_\mathsf{auth}$ receives the user's response, $D_\mathsf{auth}$ forwards the email response verbatim, including all headers, to all other discovery nodes.
    Each discovery node independently checks if (1) the challenge it generated was included in the email, (2) the DKIM signature is valid, using the public key stored in DNS for the domain of $\mathit{ID}$, and (3) the registration request was not marked as invalid at Step \ref{item:first-node-check-id}. If any of these fail, the discovery node aborts the registration process.

    \item Each discovery node that successfully completes the previous steps sends a signed \textit{confirmation message} to all other discovery nodes, which includes Alice's $\mathit{ID}$ and $\Delta$. 
    Once a discovery node has received messages from $2f$ discovery nodes for Alice's $\mathit{ID}$, it stores the link between $\mathit{ID}$ and $\Delta$ in its database.
    Waiting for confirmation from $2f$ distinct discovery nodes ensures all correct discovery nodes register the same $\Delta$, even if $D_\mathsf{auth}$ is malicious.
    Each discovery node then sends a message via Loopix to the user's contact information, confirming registration.

    \item If the user device receives at least $2f+1$ confirmations within a predetermined timeout period, the user knows that she is successfully registered in Pudding.
\end{enumerate}

If $D_\mathsf{auth}$ is malicious, it is possible with this protocol that some honest discovery nodes do not complete the registration.
This can be avoided by disseminating $(\mathit{ID}, \Delta)$ using a Byzantine reliable broadcast protocol \cite{bracha1987,cachin2011}.

\section{Evaluation}\label{sec:evaluation}

To evaluate our protocol, we compare it against our security goals (\S\ref{sec:security-analysis}), discuss its limitations (\S\ref{sec:limitations}), and present findings from implementing Pudding over Nym (\S\ref{sec:practical_evaluation}).

\subsection{Security analysis}\label{sec:security-analysis}
We begin by formalising the goals \ref{goal:unlinkability}\dots\ref{goal:membership_unobservability} from \S\ref{sec:threat_model} as security games.
We denote the adversary as \adversary and challenger as \challenger.
$U_\challengermath$ and $U_\adversarymath$ refer to users controlled by the challenger and adversary, respectively. 

\bolditem{Unlinkability game \unlinkabilitygame:}\label{item:unlinkability_security_analysis} 
The adversary \adversary controls up to $f$ discovery nodes, the user $U_\adversarymath$, the mix nodes in all but one of the Loopix network layers, and the network between all of the nodes.
The challenger \challenger controls users $U_\challengermath$, $U_0$, and $U_1$ chosen by the adversary, and generates a random bit $b \mathbin{\in_R} \{0, 1\}$. 
$U_\challengermath$ is registered to the discovery nodes with the username $\mathit{ID}_\challengermath$.
$U_{b}$ runs the discovery protocol (\lookup, \continit, and \addfriend) with input $\mathit{ID}_\challengermath$, and $U_\challengermath$ decides to respond in \continit Step~\ref{item:continit-reply-ignore}.
Finally, \adversary outputs $b'$. 
The adversary wins if $b=b'$.

\bolditem{\unlinkabilitygame security arguments:} 
    We claim that the adversary cannot win \unlinkabilitygame with a significantly higher probability than the chance of breaking sender-receiver unlinkability in the underlying Loopix network (which depends on the size of the anonymity set).
    The reason is threefold.
    (1) An adversary performing passive or active attacks on the network links between nodes or adversary-controlled mix nodes is limited by the unlinkability property of Loopix.
    (2) Controlling discovery nodes does not confer the adversary an advantage, since all communication between $U_{b}$ and those nodes is done via SURBs, which hide the identity of the recipient.
    (3) The potentially identifying information in $M_\mathsf{init}$ is encrypted with the blinded public key of $U_\challengermath$, which \adversary does not control.
    Appendix \ref{appendix:unlinkability-proof} provides a more detailed discussion.
    
\bolditem{Security against impersonation game \impersonationgame:} 
    There are three users: $U_\challengermath$, $U_\adversarymath$, and an honest user $U_\mathcal{H}$ controlled by \challenger.
    $U_\mathcal{H}$ registers using $\mathit{ID}$, which is an email address not controlled by \adversary and known to \challenger. 
    \adversary controls up to $f$ discovery nodes, any number of mix nodes and provider nodes, and the network between all of the nodes.
    \adversary chooses two messages $m_0$, $m_1$ of the same length and reveals them to \challenger.
    \challenger generates a random bit $b \mathbin{\in_R} \{0, 1\}$.
    $U_\challengermath$ runs the discovery protocol with input $\mathit{ID}$, and once \addfriend completes, $U_\challengermath$ sends a message $m_b$ encrypted with the session key $K_s$.
    Meanwhile, the adversary may inject arbitrary messages into the network.
    \adversary outputs $b'$, and wins the game if $b'=b$.

\bolditem{\impersonationgame security arguments:} 
    We claim that the adversary has a negligible advantage over a random guess.
    (1) $U_\challengermath$ only accepts one response per discovery node per \lookup, since responses are signed and linked with $U_\challengermath$'s nonce.
    Moreover, \adversary cannot modify the \lookup responses from the honest discovery nodes, since the Sphinx packet format ensures integrity \cite{danezis2009sphinx}.
    Therefore, $U_\challengermath$ will not reach the threshold of $f+1$ responses required to accept the adversary's responses, and therefore the SURB and blinded public key used by $U_\challengermath$ will be the one generated by the honest discovery nodes using the contact information of $U_\mathcal{H}$.
    (2) The adversary knows the nonce and SURBs for $U_\challengermath$, and can therefore send fake \addfriend protocol messages to $U_\challengermath$ and $U_\mathcal{H}$.
    However, due to the existential unforgeability of the key-blinded signature scheme, the adversary cannot forge a signature that $U_\challengermath$ will accept as valid for $U_\mathcal{H}$'s blinded public key (as produced by the honest discovery nodes).
    Therefore, the computation of $K_m$ at \addfriend Step~\ref{item:addfriend-alice-check} will only be performed using the $g^b$ sent by $U_\mathcal{H}$, and therefore \adversary does not learn $K_s$.
    Even though the adversary can trick $U_\mathcal{H}$ into sending the ciphertext of $m_b$ to $U_\adversarymath$ by replacing the SURB in \addfriend protocol messages, by the IND-CPA property of the symmetric cipher used to encrypt $m_b$, the adversary has only negligible advantage over guessing $b$.
    See Appendix \ref{appendix:impersonation_proof} for further details.
    
\bolditem{External identity verification game \idverificationgame:} 
    \adversary controls $f$ discovery nodes out of $3f+1$, as well as the user $U_\adversarymath$.
    \challenger controls the remaining $2f+1$ discovery nodes.
    $\mathit{ID}$ is an email address not controlled by \adversary.
    Per our threat model for \ref{goal:id_verification} (\S\ref{sec:threat_model}), the email service used for identity verification is modelled as an oracle, guaranteeing that an email originating from an email address includes a DKIM signature that corresponds to that particular email address.
    $U_\adversarymath$ runs the \register protocol for $\mathit{ID}$.
    \adversary wins the game if any of the $2f+1$ challenger-controlled discovery nodes registers $\mathit{ID}$.
    
\bolditem{\idverificationgame security arguments:} 
    We claim that the adversary has a negligible probability of winning game \idverificationgame. 
    In order to successfully complete the \register protocol for $\mathit{ID}$, \adversary would have to forge a digital signature and pass the checks in \register Step \ref{item:first-node-dkim-check} for any honest discovery node.

\bolditem{Membership unobservability game \unobservabilitygame:}
    \adversary controls an arbitrary number of users $U_{\adversarymath0}, U_{\adversarymath1}, \dotsc$, but \adversary does not have access to the internal states of any discovery or provider nodes, as per our threat model for \ref{goal:membership_unobservability} (\S\ref{sec:threat_model}).
    $\mathit{ID}_0$ and $\mathit{ID}_1$ are initially unregistered usernames known to both \adversary and \challenger.
    \challenger controls user $U_\challengermath$, generates a random bit $b \mathbin{\in_R} \{0, 1\}$, and runs the \register protocol to register $(\mathit{U_\challengermath:ID}_b,\, \Delta)$ for some arbitrary $\Delta$ not controlled by \adversary.
    \adversary then runs the discovery protocol for $\mathit{ID}_0$ and $\mathit{ID}_1$ from any adversary-controlled users arbitrarily many times, but $U_\challengermath$ does not respond to any of those contact requests.
    Finally, \adversary outputs $b'$ and wins the game if $b=b'$. 
    
\bolditem{\unobservabilitygame security arguments:}
    We claim that \adversary has a negligible advantage over a random guess.
    Since $U_\challengermath$ does not respond to $M_\mathsf{init}$, the only messages received by the adversary are the discovery node responses.
    These messages contain the searcher's nonce, a SURB, a blinded public key, and a signature.
    The nonce and signature do not depend on whether the username is registered; the SURB is computationally indistinguishable from a uniform random bit string \cite{danezis2009sphinx}; and the key-blinding scheme we use ensures that the blinded public key for real contact information is indistinguishable from that for $\Delta_\mathsf{fake}$ \cite{keyblindingsignatures}.

\subsection{Discussion and limitations}\label{sec:limitations}
\point{Email service provider}
There are two attacks that an adversary who controls a corrupted email service provider can perform.
(1) An adversary who controls the user's SMTP server, or who can intercept unencrypted SMTP traffic, can register with Pudding on behalf of any email address on that server.
(2) DKIM works at the \textit{domain level}: it does not strictly verify that the part of the email address before the `@' sign is correct, but that an email has been signed by an email server that holds the private key for a particular domain \cite{crocker2011domainkeys}. 
Therefore, if the adversary has access to the domain's signing key, the adversary can forge emails with valid DKIM signatures, which appear to be sent from any email address hosted by that domain. 

\point{Email traffic}
Even if the attacker does not control an email service provider, she can try to learn if a user has registered to Pudding with a certain email address by observing the network traffic between the user device and an email service provider, or between $D_\mathsf{auth}$ and the email service provider (e.g. for personal domains).
This attack is outside our threat model.

\point{Malicious Loopix providers}
A Loopix provider node receives messages on behalf of its users while the users are offline, and allows users to download those messages when they next connect to the provider.
This design is advantageous for mobile devices that are frequently offline, but it has the disadvantage that a malicious provider can learn the number of messages received over time by each of its users (but not their origin).
Some messaging patterns, such as receiving $3f+1$ messages in short succession, could indicate that the user has just performed a \lookup request.
By linking the timing of such message receipt patterns with the timing of \lookup requests at a colluding discovery node, an adversary can make statistical estimates of which users might be searching for each other.
This attack can be mitigated by ensuring that users receive sufficient (real or cover) traffic, and by increasing the size of the anonymity set by having more users submitting \lookup requests.

\point{Spam}
Spam is usually detected by looking at message contents.
However, end-to-end encrypted messaging apps can reduce spam by detecting high volumes of automated messages coming from the same source and directed at a high number of users \cite{jones2017whatsappspam}.
Since Pudding does not have access to such metadata, we should use other measures to prevent spam.
One strict yet effective measure can be dropping the message if the person is not in your contact address book. 
Another measure could be making spamming expensive; for example, Nym requires its users to purchase bandwidth credentials.
    
\point{Pseudonymous usernames}
It may seem that Pudding should support pseudonymous identities for users who do not want to tie a non-anonymous identifier, such as an email address, to their anonymity network presence.
However, removing external verification from registration also eliminates any chance of achieving membership unobservability.
If anybody can register any username, an adversary can check whether username \textit{ID} is registered by attempting to register \textit{ID}, and then using another device to discover \textit{ID} and send a message to it. 
If the adversary receives her own message, then \textit{ID} was not previously registered. 
If the adversary does not receive her own message, even after several retries, then \textit{ID} must have already been registered.

Consequently, in a seemingly paradoxical manner, tying anonymity network accounts to external identities such as email addresses provides \emph{better} privacy than using short pseudonymous identities, in terms of hiding membership of the network.
We have therefore chosen not to include pseudonymous usernames as a registration option in Pudding.
However, we do not require users to register to discover others or send messages.
Only one party in a conversation needs to register with Pudding in order to initiate a two-way conversation.

\point{Online / offline patterns}
Loopix hides communication patterns, however the provider nodes can still determine whether a user device is online or offline.
If device pairs transition between online and offline states in a correlated fashion, it can potentially be used to infer that two users are communicating. 
However, this type of intersection attack is not unique to Loopix, but rather a concern in all anonymity networks \cite{danezis2004statistical}. 
This vulnerability is also acknowledged by other private user discovery protocols, as highlighted in Alpenhorn \cite{lazar2016alpenhorn} and UDM \cite{chaum2021udm}.

\point{Usability}
In this work, we base Pudding’s usability traits upon previous work.
We argue that enabling user discovery through usernames increases usability, because previous research shows that users are not good at generating and exchanging information needed to establish communication over an encrypted channel \cite{whitten1999johnny, ruoti2015johnny, sheng2006johnny}.
Using email addresses as usernames increases usability, based on the fact that all popular encrypted messaging apps allow users to find their friends through identifiers that already exist in their address books.
The requirement to allow user discovery without leaving the network draws from research showing that authentication ceremonies of encrypted messaging apps have poor usability \cite{vaziripour2017you, herzberg2016can}. 
In such ceremonies, users share public keys or verification codes manually with each other via QR code, near-field communication (NFC), or through speech \cite{unger2015sok}. 
Although during registration to Pudding the user has to leave the anonymity network to respond to a verification email, this only happens once for each username registration, and the user typically does not need to take any action other than hitting the reply button.
Further work is required to measure the specific usability improvement that our protocol presents. 

\subsection{Performance evaluation}\label{sec:performance-analysis}
\label{sec:practical_evaluation}

We have implemented a prototype of Pudding on Nym to demonstrate that our protocol is practical.
The prototype is implemented in 3000 lines of Rust code using the Nym SDK\footnote{\url{https://nymtech.net/docs/sdk/rust.html}}.
We also modified the SDK and the \texttt{sphinx-packet} crate to expose and modify internal methods that allow us to create deterministic SURBs and access the user's secret keys.
The changes are contained in a \texttt{.patch} file with less than 2000 lines.
Importantly, no modifications of Nym infrastructure (mix nodes, provider nodes, or protocols) are required.
As such, all our experiments run on the general main net and share the same anonymity set as existing Nym users.
We have chosen Nym because it is an operational anonymity network running on hundreds of nodes, and it does not currently provide a user discovery mechanism.
All source code and SDK modifications are available in our open-source repository under an MIT license: \url{https://github.com/ckocaogullar/pudding-protocol}.

The Nym SDK manages SURBs when sending messages to other recipients using their Nym address.
When sending a message from a user $U_\mathsf{sender}$ to another user $U_\mathsf{recipient}$, the Nym client transparently includes SURBs so that $U_\mathsf{recipient}$ can reply without knowing $U_\mathsf{sender}$'s address.
As messages are exchanged, $U_\mathsf{recipient}$ requests more SURBs from $U_\mathsf{sender}$ when her stock runs low.
This mechanism is used in our protocol when the user communicates with the discovery servers.
However, when using our deterministic SURBs (\S\ref{sec:deterministic-surb-gen}) both parties are anonymous towards each other and we have to make sure ourselves that the responses include sufficiently many SURBs.

In our evaluation, all clients run on a machine with an 8-core CPU (Intel Xeon Skylake) in a commercial data center.
During an initialisation phase, the test program creates ephemeral clients with randomly chosen gateways for the users and discovery servers.
All clients execute in a multi-threaded asynchronous runtime inside the same process.
However, they do not share any information out-of-band once started.
The DKIM verification step is simulated by another Nym client that replies with a signature immediately, effectively excluding the time taken by the user to receive and reply to an email. 

In this evaluation, we are interested in the latency distributions for the Pudding sub-protocols. 
Specifically, we are interested in four metrics: latency of the \register protocol (\S\ref{sec:registration}), latency of the \lookup protocol (\S\ref{sec:lookup}), and the latency for the combination of the \continit and \addfriend protocols in both the anonymous and non-anonymous variant.
Since \lookup is required for \continit, these protocols are executed as part of the same scenario.
Therefore, we execute three different scenarios: a registration scenario, an anonymous discovery (\lookup+\continit+\addfriend) scenario, and a non-anonymous discovery scenario.
We run each scenario for 10\,minutes and repeat it 6 times to account for outliers, such as unlucky choice of slow gateway nodes.
We run configurations with 20 client nodes and a varying number of $n=\{4, 7, 10\}$ discovery nodes.

For the \register scenario, each client starts a new registration attempt every 30\,seconds after an initial random pause.
Similarly, for the other scenarios every client sends a \lookup request for a randomly chosen other user every 30\,seconds and proceeds then with the \continit and \addfriend steps.
For the discovery scenarios, all clients are pre-registered with the discovery nodes such that the results are not influenced by pending registration operations.
We stop the clock for the \register operation once the user receives $2f+1$ confirmations, and for the \lookup operation once the searcher receives $f+1$ confirmations.
The combination of \continit + \addfriend is measured from the time when the searcher sends her first message to receiving the confirmation from the searchee, at which point the searcher and searchee can send messages to each other.

We have verified that the number of users (up to the maximum that our test bed can support) has negligible impact on the latency distributions.
We also noted occasional instances where some messages were lost or nodes experienced temporary disconnections from the Nym network.
Since these cases are automatically covered by the fault-tolerance of our protocol, we observed little difference between executions with failed nodes and those without.

\begin{figure}
    \centering
    \includegraphics[width=\columnwidth]{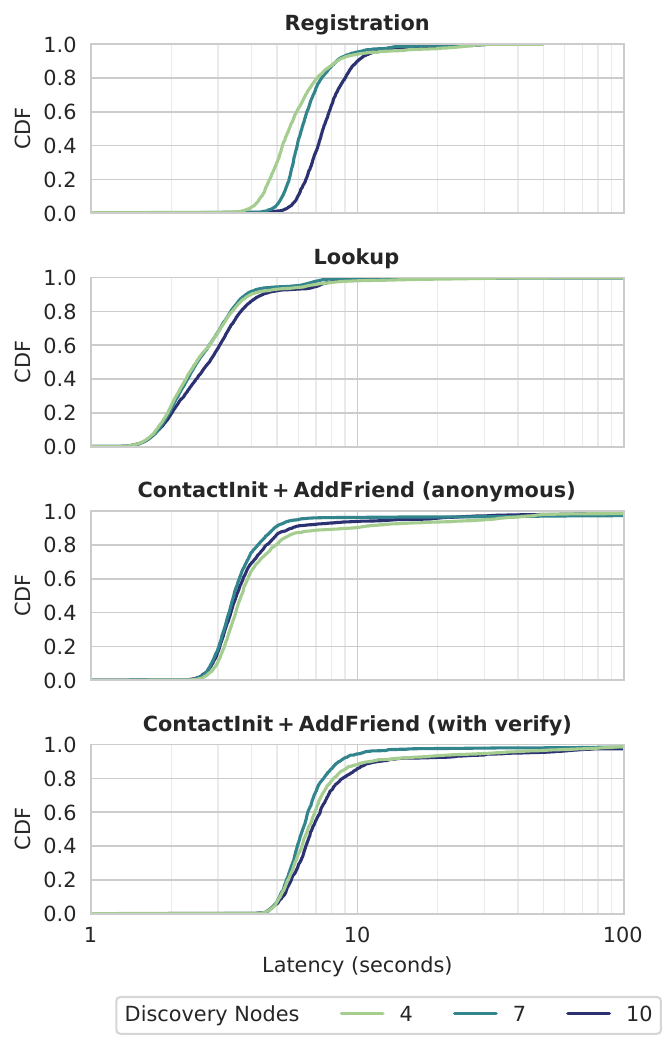}
    \caption{Latency of the Pudding sub-protocols represented as CDFs for different numbers of discovery nodes. 
    Note that the x-Axis is log-scaled.
    See also Table~\ref{tab:latencies} in Appendix~\ref{sec:appendix_performance}.}
    \label{fig:latencies}
\end{figure}

\hfill \break

\textbf{Registration and discovery over Nym are fast.}
Our results are presented as cumulative distributions in Figure~\ref{fig:latencies}, while detailed numerical values can be found in Appendix \ref{sec:appendix_performance}.
All operations have mean latencies below \upperboundtotalmean seconds.
We therefore believe that both registration and discovery latency are suitable for real-world usage.

For this paragraph, we consider the scenario with $n=10$ discovery servers.
The mean latency for \register is \registermean seconds ($p_{90}=$ \registerpninety\,s).
These numbers are lower for the $n=4$ configuration.
This is because, as $n$ increases, the client has to wait for a larger number of replies which increases the chance of having to wait for a message with large mix node delays.
The mean \lookup latency is \lookupmean\,seconds ($p_{90}=$ \lookuppninety\,s).
It is similar for other $n$ as there are no ``synchronization barriers'', where a single discovery node has to collect message from the others.
The latency for the combined \continit and \addfriend step differs between the anonymous and non-anonymous variants.
For the former, the mean latency is \anonmean\,seconds ($p_{90}=$ \anonpninety\,s) while the non-anonymous variant has a mean latency of \nonanonmean\,seconds ($p_{90}=$ \nonanonpninety\,s).
The non-anonymous variant is expected to be slower, because the searchee has to perform an extra communication round with the discovery servers.

\textbf{Pudding's performance is influenced by the anonymity network's throughput limits.}
Under high load, the latencies can exhibit significant outliers.
This is because Nym clients send data at a fixed rate (adding cover traffic if necessary) to prevent traffic analysis.
Consequently, sending more messages increases the expected queueing delay.
This effect is especially pronounced for the discovery requests, where the responses from the discovery nodes exceed the size limit of a single payload in Nym.
As a result, each response to the searcher must be split up into multiple Nym packets (in our implementation we need up to ten).

In a real deployment, we anticipate that the queuing delay on user devices will be small, because registration and discovery requests are sent rarely -- occurring only once during app installation for registration and once during initial communication with a friend for discovery.
However, the discovery nodes may need to handle a high rate of requests and responses. 
This can be addressed by running multiple instances of the Nym client at each discovery node, and load-balancing messages across them.
For this, the list of discovery nodes may contain multiple Nym identifiers for each, and clients randomly pick one when sending a message to that node.

\section{Related work}\label{sec:related-work}

We know of six systems that provide user discovery while hiding who is communicating with whom: Alpenhorn \cite{lazar2016alpenhorn}, UDM \cite{chaum2021udm}, PROUD \cite{papadopoulos2018s}, OUTOPIA \cite{papadopoulos2022outopia}, Arke \cite{mohnblatt2023arke}, and Apres \cite{laurie2004apres}.

\emph{Alpenhorn} \cite{lazar2016alpenhorn} is based on the Vuvuzela \cite{van2015vuvuzela} anonymity network, and therefore provides a high level of metadata privacy. 
Alpenhorn has a round-based system that requires users to periodically download large inboxes rather than accessing individual messages on demand. 
As a result, this scheme has high bandwidth costs, which is especially challenging for devices relying on cellular networks.
For instance, with one million users, each user device needs to download a 7.5 MB mailbox, which exceeds the allowance of many cellular data plans \cite{lazar2016alpenhorn}.
The alternative of delaying mailbox downloads until WiFi is available can increase the lookup latency. 
Furthermore, if a user device fails to participate in a round where a lookup request was initiated for that user, the request will be lost.
Moreover, Alpenhorn cannot remain operational if a single server crashes, goes offline or otherwise does not follow the protocol.
Pudding meanwhile, is fault-tolerant and fast; it works by exchanging small messages rather than large mailboxes.

\emph{UDM} \cite{chaum2021udm} aims to hide contact relationships and protect users' external public identities from being linked to their private network identities. 
Compared to Pudding, UDM does not provide \textit{membership unobservability} (\ref{goal:membership_unobservability}) or \textit{external identity verification} (\ref{goal:id_verification}). 
In addition, UDM does not prevent an adversary from deducing contact relationships by passively observing network patterns. 
Lastly, UDM does not provide fault tolerance by default, and discusses relevant measures as extensions to the current model. 

\emph{PROUD} \cite{papadopoulos2018s} is a DNS-based private user discovery system that does not provide traffic analysis protection and requires users to exchange public keys out-of-band. 
Similarly, \emph{OUTOPIA} \cite{papadopoulos2022outopia} requires users to exchange out-of-band information such as public keys and ``long hard-to-guess names'', and is therefore not a usable discovery mechanism. 
Neither of these schemes meet our usability goals.

\emph{Arke} \cite{mohnblatt2023arke} is a contact discovery protocol based on an unlinkable handshake mechanism, built atop an identity-based non-interactive key exchange. 
While it provides Byzantine fault tolerance like Pudding, Arke does not offer inherent unlinkability against passive adversaries observing network traffic.
Also notably, Arke requires mutual interest for contact initiation, making it unsuitable for scenarios where one party seeks to contact another unilaterally, such as a whistleblower reaching out to a journalist.

\emph{Apres} \cite{laurie2004apres} allows users to find and message each other privately, providing unlikability and online unobservability.
However, Apres requires each pair of users to communicate out-of-band and determine a pairwise introduction ID, which must remain secret and therefore also be hard-to-guess.
In contrast, Pudding users have a single non-secret ID.

Private discovery of network entities is relevant for many types of anonymous and other communication networks. 
Below, we detail some prominent techniques and systems used for various private discovery purposes:

\emph{Private Set Intersection (PSI)} protocols allow two or more parties to find the intersection of sets without learning anything else about each other's sets, except maybe their sizes \cite{demmler2018pir}. 
PSI is a suitable solution for privately determining the intersection of the contacts in a user's address book and the users registered in the communication system \cite{kales2019mobile}. 
However, PSI is not directly applicable to retrieving contact information.

\emph{Private Information Retrieval (PIR)} protocols allow querying items from a server without revealing the query to the server \cite{pung}.
Demmler et al. used a technique combining PIR with PSI to privately find which users in someone's contact address book are registered in a messaging system \cite{demmler2018pir}. 
However, this technique does not achieve \textit{membership unobservability} (\ref{goal:membership_unobservability}).
DP5 \cite{borisov2015dp5} also aims to solve a related but different problem: privately learning the online status of one's contacts. 
DP5 also does not provide \textit{membership unobservability} (\ref{goal:membership_unobservability}), and it requires users to exchange private keys out-of-band.

\emph{Trusted Execution Environments (TEEs)} are used for protecting the confidentiality of data \textit{in-use}. 
In the user discovery context, Signal uses Intel Software Guard Extensions (SGX) to privately find the intersection of the user's address book and the users registered to the messenger service \cite{signalcontactdiscovery}. 
TEEs can also be integrated into Pudding discovery nodes, to reduce the level of trust associated with discovery node providers.

\emph{Private DNS solutions} including DNSCrypt \cite{dnscrypt}, DNSCurve \cite{dnscurve}, Confidential DNS \cite{confidential-dns} and Private DNS \cite{private-dns}, focus on protecting DNS queries and responses from being read or modified by unauthorised parties. 
These mechanisms are not suitable for private user discovery in anonymity networks, since they do not prevent the DNS servers from linking users with their queries. 
More recent approaches like Oblivious DNS \cite{schmitt2019oblivious}, DNS for Tor \cite{dnsfortor}, and distributed DNS \cite{hoang2020distributeddoh} are not resistant against global passive adversaries.

\emph{Distributed Hash Tables (DHTs)}\label{background:dht} allow peers in the system to search for data objects using the keys associated with them. 
As with DNS, most DHT implementations do not provide protection against linking users with queries. 
The ones that aim to provide this property, including Salsa \cite{salsa}, AP3 \cite{ap3}, NISAN \cite{nisan}, Torsk \cite{torsk} are susceptible to information leakage, as illustrated by numerous studies \cite{nisan, p2p-anonymous-lookup-1, p2p-anonymous-lookup-2, danezis2008bridging}. 

\section{Conclusions}
Anonymity networks that allow messaging, such as Loopix \cite{loopix} and Nym, currently lack a usable and privacy-preserving user discovery mechanism. 
Previous research suggests \cite{whitten1999johnny} this might impede widespread adoption. 
In this work, we have described Pudding, a novel private user discovery protocol that hides contact relationships, prevents impersonation, and provides fault-tolerance. 
Our protocol also provides membership unobservability against malicious users using a new deterministic single-use reply block (SURB) generation mechanism. 
Users register to Pudding with their email address.
We verify email ownership of through a mechanism based on DomainKey Identified Mails (DKIM). 
A messaging app can then automatically search for the friends of the user by using the email addresses of friends stored in her address book, confident that the recipient is in control of the associated email inbox. 
Pudding is an application-layer protocol; it can be implemented as an overlay network on top of any anonymity network that supports SURBs. 
We have implemented our protocol, and evaluated its performance on Nym anonymity network. 
Our evaluation shows that Pudding is scalable, practical, and lightweight, achieving mean latencies of under \lookupupperboundmean\,seconds for lookup, \nonanonupperboundmean\,seconds for initiating contact, and \registerupperboundmean\,seconds for registration---numbers significantly influenced by the delays in the underlying anonymity network.

\section*{Acknowledgements}
We would like to thank Alberto Sonnino and the anonymous reviewers for their valuable feedback.
Special thanks to Furkan Saygın Şener for his creativity in naming the protocol.
Ceren Kocaoğullar is supported by King's College, Cambridge and the Cambridge Trust.
She was supported by Nokia Bell Labs during the initial stages of this research project.
Daniel Hugenroth is supported by Nokia Bell Labs.
Martin Kleppmann is funded by the Volkswagen Foundation and crowdfunding supporters including Rohit Kulshreshtha, Mintter, and SoftwareMill.

\bibliographystyle{IEEEtran}
\bibliography{main}

\begin{thebibliography}{10}
\providecommand{\url}[1]{#1}
\csname url@samestyle\endcsname
\providecommand{\newblock}{\relax}
\providecommand{\bibinfo}[2]{#2}
\providecommand{\BIBentrySTDinterwordspacing}{\spaceskip=0pt\relax}
\providecommand{\BIBentryALTinterwordstretchfactor}{4}
\providecommand{\BIBentryALTinterwordspacing}{\spaceskip=\fontdimen2\font plus
\BIBentryALTinterwordstretchfactor\fontdimen3\font minus \fontdimen4\font\relax}
\providecommand{\BIBforeignlanguage}[2]{{%
\expandafter\ifx\csname l@#1\endcsname\relax
\typeout{** WARNING: IEEEtran.bst: No hyphenation pattern has been}%
\typeout{** loaded for the language `#1'. Using the pattern for}%
\typeout{** the default language instead.}%
\else
\language=\csname l@#1\endcsname
\fi
#2}}
\providecommand{\BIBdecl}{\relax}
\BIBdecl

\bibitem{whitten1999johnny}
A.~Whitten and J.~D. Tygar, ``Why {Johnny} can't encrypt: A usability evaluation of {PGP} 5.0.'' in \emph{{USENIX Security}}, 1999.

\bibitem{ukrainecellphones}
\BIBentryALTinterwordspacing
M.~Srivastava, M.~Murgia, and H.~Murphy, ``The secret mission to bolster {Ukraine}'s cyber defences ahead of {Russia}'s invasion,'' Financial Times, Mar. 2022, (accessed 2021-08-27). [Online]. Available: \url{https://perma.cc/3YZC-863H}
\BIBentrySTDinterwordspacing

\bibitem{strohmeier2018real}
M.~Strohmeier, M.~Smith, V.~Lenders, and I.~Martinovic, ``The real first class? {I}nferring confidential corporate mergers and government relations from air traffic communication,'' in \emph{EuroS\&P}, 2018.

\bibitem{fitbitheatmap}
\BIBentryALTinterwordspacing
``Fitness app {Strava} lights up staff at military bases,'' BBC News, Jan. 2018, (accessed 2021-08-27). [Online]. Available: \url{https://perma.cc/ZWS2-5VTC}
\BIBentrySTDinterwordspacing

\bibitem{bellingcatnavalny}
\BIBentryALTinterwordspacing
A.~Toler, ``Hunting the hunters: How we identified {Navalny}'s {FSB} stalkers,'' Bellingcat.com, Dec. 2020, (accessed 2021-08-27). [Online]. Available: \url{https://perma.cc/25JU-V4A4}
\BIBentrySTDinterwordspacing

\bibitem{signalsealedsender}
\BIBentryALTinterwordspacing
J.~Lund, ``Technology preview: Sealed sender for {Signal},'' Signal blog, Oct. 2018, (accessed 2021-05-09). [Online]. Available: \url{https://perma.cc/6HC9-Y44E}
\BIBentrySTDinterwordspacing

\bibitem{loopix}
A.~M. Piotrowska, J.~Hayes, T.~Elahi, S.~Meiser, and G.~Danezis, ``The {Loopix} anonymity system,'' in \emph{{USENIX Security}}, 2017.

\bibitem{van2015vuvuzela}
J.~Van Den~Hooff, D.~Lazar, M.~Zaharia, and N.~Zeldovich, ``Vuvuzela: Scalable private messaging resistant to traffic analysis,'' in \emph{ACM SOSP}, 2015.

\bibitem{talek}
R.~Cheng, W.~Scott, E.~Masserova, I.~Zhang, V.~Goyal, T.~Anderson, A.~Krishnamurthy, and B.~Parno, ``Talek: Private group messaging with hidden access patterns,'' in \emph{{Annual Computer Security Applications Conference (ACSAC)}}, 2020.

\bibitem{pung}
S.~Angel and S.~Setty, ``{Unobservable communication over fully untrusted infrastructure},'' in \emph{USENIX OSDI}, 2016.

\bibitem{signalcontactdiscovery}
\BIBentryALTinterwordspacing
M.~Marlinspike, ``Technology preview: Private contact discovery for {Signal},'' Signal blog, Sep. 2017, (accessed 2021-09-02). [Online]. Available: \url{https://perma.cc/N3GF-8WD6}
\BIBentrySTDinterwordspacing

\bibitem{whatsapprivacypolicy}
\BIBentryALTinterwordspacing
{WhatsApp Ltd}, ``{WhatsApp} privacy policy,'' (accessed 2021-09-02). [Online]. Available: \url{https://perma.cc/CC39-2CFT}
\BIBentrySTDinterwordspacing

\bibitem{hagen2021all}
C.~Hagen, C.~Weinert, C.~Sendner, A.~Dmitrienko, and T.~Schneider, ``All the numbers are {US}: Large-scale abuse of contact discovery in mobile messengers,'' in \emph{NDSS}, 2021.

\bibitem{lazar2016alpenhorn}
D.~Lazar and N.~Zeldovich, ``Alpenhorn: Bootstrapping secure communication without leaking metadata,'' in \emph{USENIX OSDI}, 2016.

\bibitem{chaum2021udm}
D.~Chaum, M.~Yaksetig, A.~T. Sherman, and J.~de~Ruiter, ``{UDM}: Private user discovery with minimal information disclosure,'' \emph{Cryptologia}, vol.~46, no.~4, pp. 347--379, 2022.

\bibitem{onionservices}
\BIBentryALTinterwordspacing
``Tor rendezvous specification - version 3,'' Tor Project, (accessed 2021-09-02). [Online]. Available: \url{https://gitweb.torproject.org/torspec.git/tree/rend-spec-v3.txt}
\BIBentrySTDinterwordspacing

\bibitem{tor}
R.~Dingledine, N.~Mathewson, and P.~Syverson, ``Tor: The second-generation onion router,'' Naval Research Lab Washington DC, Tech. Rep., 2004.

\bibitem{mixminion}
G.~Danezis, R.~Dingledine, and N.~Mathewson, ``{Mixminion: Design of a type III anonymous remailer protocol},'' in \emph{{IEEE S\&P}}, 2003.

\bibitem{danezis2009sphinx}
G.~Danezis and I.~Goldberg, ``Sphinx: A compact and provably secure mix format,'' in \emph{{IEEE S\&P}}, 2009.

\bibitem{unger2015sok}
N.~Unger, S.~Dechand, J.~Bonneau, S.~Fahl, H.~Perl, I.~Goldberg, and M.~Smith, ``{SoK}: secure messaging,'' in \emph{IEEE S\&P}, 2015.

\bibitem{chaum1981untraceable}
D.~L. Chaum, ``Untraceable electronic mail, return addresses, and digital pseudonyms,'' \emph{Communications of the ACM}, vol.~24, no.~2, pp. 84--90, 1981.

\bibitem{backes2013anoa}
M.~Backes, A.~Kate, P.~Manoharan, S.~Meiser, and E.~Mohammadi, ``{AnoA}: A framework for analyzing anonymous communication protocols,'' in \emph{IEEE Computer Security Foundations Symposium}, 2013.

\bibitem{Paverd2014}
\BIBentryALTinterwordspacing
A.~Paverd, A.~Martin, and I.~Brown, ``Modelling and automatically analysing privacy properties for honest-but-curious adversaries,'' University of Oxford, Tech. Rep., 2014. [Online]. Available: \url{https://ajpaverd.org/publications/casper-privacy-report.pdf}
\BIBentrySTDinterwordspacing

\bibitem{Lamport:1982fr}
L.~Lamport, R.~Shostak, and M.~Pease, ``The {Byzantine} generals problem,'' \emph{ACM Transactions on Programming Languages and Systems}, vol.~4, no.~3, pp. 382--401, Jul. 1982.

\bibitem{hashtocurve}
\BIBentryALTinterwordspacing
A.~Faz-Hernandez, S.~Scott, N.~Sullivan, R.~S. Wahby, and C.~A. Wood, ``Hashing to elliptic curves,'' IETF Internet-Draft, 2023. [Online]. Available: \url{https://datatracker.ietf.org/doc/draft-irtf-cfrg-hash-to-curve/}
\BIBentrySTDinterwordspacing

\bibitem{keyblindingsignatures}
\BIBentryALTinterwordspacing
F.~Denis, E.~Eaton, T.~Lepoint, and C.~A. Wood, ``Key blinding for signature schemes,'' IETF Internet-Draft, 2023. [Online]. Available: \url{https://datatracker.ietf.org/doc/draft-irtf-cfrg-signature-key-blinding/}
\BIBentrySTDinterwordspacing

\bibitem{eaton2021post}
E.~Eaton, D.~Stebila, and R.~Stracovsky, ``Post-quantum key-blinding for authentication in anonymity networks,'' in \emph{LATINCRYPT}, 2021.

\bibitem{hopperproving}
N.~Hopper, ``Proving security of {Tor’s} hidden service identity blinding protocol,'' Tor project, Tech. Rep., 2013.

\bibitem{chaum1983blind}
D.~Chaum, ``Blind signatures for untraceable payments,'' in \emph{CRYPTO}, 1983.

\bibitem{krawczyk2003sigma}
H.~Krawczyk, ``{SIGMA}: The {`SIGn-and-MAc'} approach to authenticated {Diffie-Hellman} and its use in the {IKE}-protocols.'' in \emph{CRYPTO}, 2003.

\bibitem{canetti2002sigma}
R.~Canetti and H.~Krawczyk, ``Security analysis of {IKE}'s signature-based key-exchange protocol,'' in \emph{CRYPTO}, 2002.

\bibitem{leiba2007domainkeys}
B.~Leiba and J.~Fenton, ``{DomainKeys Identified Mail} {(DKIM)}: Using digital signatures for domain verification.'' in \emph{{Conference on Email and Anti-Spam}}, 2007.

\bibitem{crocker2011domainkeys}
D.~Crocker, T.~Hansen, and M.~Kucherawy, ``{DomainKeys Identified Mail} {(DKIM)} signatures,'' RFC6376, 2011.

\bibitem{bracha1987}
G.~Bracha, ``Asynchronous {Byzantine} agreement protocols,'' \emph{Information and Computation}, vol.~75, no.~2, pp. 130--143, 1987.

\bibitem{cachin2011}
C.~Cachin, R.~Guerraoui, and L.~Rodrigues, \emph{Introduction to Reliable and Secure Distributed Programming}, 2nd~ed.\hskip 1em plus 0.5em minus 0.4em\relax Springer, 2011.

\bibitem{jones2017whatsappspam}
M.~Jones, ``How {WhatsApp} reduced spam while launching end-to-end encryption,'' in \emph{USENIX Enigma}, 2017.

\bibitem{danezis2004statistical}
G.~Danezis and A.~Serjantov, ``Statistical disclosure or intersection attacks on anonymity systems,'' in \emph{International Workshop on Information Hiding}, 2004.

\bibitem{ruoti2015johnny}
S.~Ruoti, J.~Andersen, D.~Zappala, and K.~Seamons, ``Why {Johnny} still, still can't encrypt: Evaluating the usability of a modern {PGP} client,'' \emph{arXiv preprint arXiv:1510.08555}, 2015.

\bibitem{sheng2006johnny}
S.~Sheng, L.~Broderick, C.~A. Koranda, and J.~J. Hyland, ``Why {Johnny} still can’t encrypt: evaluating the usability of email encryption software,'' in \emph{ACM SOUPS}, 2006.

\bibitem{vaziripour2017you}
E.~Vaziripour, J.~Wu, M.~O'Neill, J.~Whitehead, S.~Heidbrink, K.~Seamons, and D.~Zappala, ``Is that you, {Alice}? {A} usability study of the authentication ceremony of secure messaging applications,'' in \emph{SOUPS}, 2017.

\bibitem{herzberg2016can}
A.~Herzberg and H.~Leibowitz, ``Can {Johnny} finally encrypt? {E}valuating {E2E}-encryption in popular {IM} applications,'' in \emph{STAST}, 2016, pp. 17--28.

\bibitem{papadopoulos2018s}
P.~Papadopoulos, A.~A. Chariton, E.~Athanasopoulos, and E.~P. Markatos, ``Where's {Wally}? {H}ow to privately discover your friends on the {Internet},'' in \emph{ACM ASIACCS}, 2018.

\bibitem{papadopoulos2022outopia}
P.~Papadopoulos, A.~A. Chariton, M.~Pachilakis, and E.~P. Markatos, ``{OUTOPIA}: private user discovery on the internet,'' in \emph{EuroSec}, 2022.

\bibitem{mohnblatt2023arke}
\BIBentryALTinterwordspacing
N.~Mohnblatt, A.~Sonnino, K.~Gurkan, and P.~Jovanovic, ``Arke: Scalable and {Byzantine} fault tolerant privacy-preserving contact discovery,'' Cryptology ePrint Archive, 2023. [Online]. Available: \url{https://eprint.iacr.org/2023/1218}
\BIBentrySTDinterwordspacing

\bibitem{laurie2004apres}
B.~Laurie, ``Apres-a system for anonymous presence,'' Tech. Rep., 2004.

\bibitem{demmler2018pir}
D.~Demmler, P.~Rindal, M.~Rosulek, and N.~Trieu, ``{PIR-PSI}: scaling private contact discovery,'' in \emph{{PETS}}, 2018.

\bibitem{kales2019mobile}
D.~Kales, C.~Rechberger, T.~Schneider, M.~Senker, and C.~Weinert, ``Mobile private contact discovery at scale,'' in \emph{{USENIX Security}}, 2019.

\bibitem{borisov2015dp5}
N.~Borisov, G.~Danezis, and I.~Goldberg, ``{DP5}: A private presence service.'' in \emph{PoPETS}, 2015.

\bibitem{dnscrypt}
\BIBentryALTinterwordspacing
{OpenDNS}, ``{DNSCrypt},'' accessed 2021-01-07. [Online]. Available: \url{https://perma.cc/UDW6-7HYA}
\BIBentrySTDinterwordspacing

\bibitem{dnscurve}
\BIBentryALTinterwordspacing
M.~Dempsky, ``{DNSCurve}: Link-level security for the {Domain Name System},'' IETF Internet-Draft, 2009. [Online]. Available: \url{https://tools.ietf.org/id/draft-dempsky-dnscurve-00.html}
\BIBentrySTDinterwordspacing

\bibitem{confidential-dns}
\BIBentryALTinterwordspacing
W.~Wijngaards and G.~Wiley, ``Confidential {DNS},'' IETF Internet-Draft, 2015. [Online]. Available: \url{https://tools.ietf.org/html/draft-wijngaards-dnsop-confidentialdns-03}
\BIBentrySTDinterwordspacing

\bibitem{private-dns}
\BIBentryALTinterwordspacing
P.~Hallam-Baker, ``Private {DNS},'' IETF Internet-Draft, 2014. [Online]. Available: \url{https://tools.ietf.org/html/draft-hallambaker-privatedns-00}
\BIBentrySTDinterwordspacing

\bibitem{schmitt2019oblivious}
P.~Schmitt, A.~Edmundson, A.~Mankin, and N.~Feamster, ``Oblivious {DNS}: Practical privacy for {DNS} queries,'' in \emph{{PETS}}, 2019.

\bibitem{dnsfortor}
\BIBentryALTinterwordspacing
M.~Sayrafi, ``Introducing {DNS} resolver for {Tor},'' The {Cloudflare} Blog, May 2018, (accessed 2021-05-14). [Online]. Available: \url{https://perma.cc/NJY9-GDMD}
\BIBentrySTDinterwordspacing

\bibitem{hoang2020distributeddoh}
N.~P. Hoang, I.~Lin, S.~Ghavamnia, and M.~Polychronakis, ``K-resolver: towards decentralizing encrypted {DNS} resolution,'' \emph{arXiv preprint arXiv:2001.08901}, 2020.

\bibitem{salsa}
A.~Nambiar and M.~Wright, ``Salsa: a structured approach to large-scale anonymity,'' in \emph{{ACM CCS}}, 2006.

\bibitem{ap3}
A.~Mislove, G.~Oberoi, A.~Post, C.~Reis, P.~Druschel, and D.~S. Wallach, ``{AP3}: Cooperative, decentralized anonymous communication,'' in \emph{{ACM SIGOPS European Workshop}}, 2004.

\bibitem{nisan}
A.~Panchenko, S.~Richter, and A.~Rache, ``Nisan: network information service for anonymization networks,'' in \emph{{ACM CCS}}, 2009.

\bibitem{torsk}
J.~McLachlan, A.~Tran, N.~Hopper, and Y.~Kim, ``{Scalable onion routing with Torsk},'' in \emph{{ACM CCS}}, 2009.

\bibitem{p2p-anonymous-lookup-1}
P.~Mittal and N.~Borisov, ``{Information leaks in structured peer-to-peer anonymous communication systems},'' \emph{{ACM Transactions on Information and System Security}}, no.~1, pp. 1--28, 2012.

\bibitem{p2p-anonymous-lookup-2}
Q.~Wang, P.~Mittal, and N.~Borisov, ``In search of an anonymous and secure lookup: attacks on structured peer-to-peer anonymous communication systems,'' in \emph{{ACM CCS}}, 2010.

\bibitem{danezis2008bridging}
G.~Danezis and P.~Syverson, ``Bridging and fingerprinting: Epistemic attacks on route selection,'' in \emph{{PETS}}, 2008.

\end{thebibliography}

\newpage\appendices
\section{Performance measurement data}
\label{sec:appendix_performance}

We provide more detailed data for our evaluation measurements from \S\ref{sec:practical_evaluation} in Table~\ref{tab:latencies}.
For each configuration we provide the mean latency and the 50-th, 90-th and 95-th percentile of the latency distribution.

\begin{table}[hbt!]
\centering
\begin{tabular}{S | S[table-format=3.2] S[table-format=3.2] S[table-format=3.2] S[table-format=3.2]} \toprule
$\mbox{\# Discovery}$ & \multicolumn{4}{c}{Latency [s]}\\
$\mbox{Nodes}$  & $\mbox{mean}$ & $p_{50}$ & $p_{90}$ & $p_{95}$\\
\midrule
\multicolumn{5}{c}{\textit{Registration}}\\
\midrule
4 & 6.48 & 5.54 & 8.42 & 11.01 \\
7 & 6.74 & 6.18 & 8.35 & 9.74 \\
10 & 8.04 & 7.44 & 10.02 & 11.37 \\
\midrule
\multicolumn{5}{c}{\textit{Lookup}}\\
\midrule
4 & 3.26 & 2.48 & 3.97 & 6.45 \\
7 & 3.00 & 2.51 & 3.81 & 5.73 \\
10 & 3.44 & 2.77 & 4.41 & 7.08 \\
\midrule
\multicolumn{5}{c}{\textit{ContactInit + AddFriend (anonymous)}}\\
\midrule
4 & 8.77 & 3.66 & 9.47 & 30.44 \\
7 & 10.58 & 3.43 & 4.87 & 6.01 \\
10 & 8.21 & 3.51 & 5.70 & 14.21 \\
\midrule
\multicolumn{5}{c}{\textit{ContactInit + AddFriend (with verify)}}\\
\midrule
4 & 11.51 & 6.52 & 11.16 & 32.06 \\
7 & 10.00 & 6.23 & 8.55 & 10.35 \\
10 & 12.34 & 6.75 & 11.86 & 38.07 \\
\bottomrule
\end{tabular}
\caption{Registration and discovery requests latency over Nym for different numbers of discovery nodes.}
\label{tab:latencies}
\end{table}

\section{Security proof sketches}

\subsection{Unlinkability (G1)}\label{appendix:unlinkability-proof}

    \begin{lemma}\label{lemma:network_traffic}
    The adversary's advantage in winning \unlinkabilitygame from observing or performing active attacks on the network links between nodes, or from controlling mix nodes, can be reduced to the probability of breaking sender-receiver unlinkability in the underlying Loopix network, provided that at least one mix node on the path of every message is honest.
    \end{lemma}
    
    All messages exchanged between the challenger's users and the adversary-controlled nodes are sent via Loopix \cite{loopix}, which provides \textit{sender-receiver third-party unlinkability} under a threat model that includes a global passive adversary (GPA), and protection against $(n-1)$ active attacks.
    Analytical and empirical analysis of this unlinkability property of Loopix shows that given appropriate parameters are chosen for the network (i.e.\ at least 3 mix node layers, message sending rate to delay parameter ratio $\lambda/\mu\ge2$), and assuming that there is a sufficient number of active Loopix users who are not controlled by the adversary to provide a large enough anonymity set, it is very unlikely that the adversary can identify which pairs of Loopix users are communicating by observing all traffic within the network and also controlling a portion of corrupt mix nodes \cite{loopix}.
    If the adversary were to control all mix nodes on the path of a message between $U_b$ and $U_\challengermath$, she would be able to trace the message and thereby link sender and receiver; however, assuming at least one honest mix node on the path prevents this attack.
    Moreover, since the Sphinx packet format \cite{danezis2009sphinx} used by Loopix ensures integrity, any packets that are modified by the adversary are simply dropped, which prevents an active network adversary from modifying the responses from discovery nodes to a user.

    \begin{lemma}\label{lemma:discovery_response}
    Assume the adversary controls up to $f$ discovery nodes, any number of mix nodes, and the network between the nodes.
    When an honest user $U_\challengermath$ performs a \lookup, the SURB and blinded public key she accepts at Step \ref{item:lookup-receive-surbs} of \lookup are those generated by the honest discovery nodes.
    \end{lemma}

    $U_\challengermath$ waits to receive identical responses from at least $f+1$ distinct discovery nodes before accepting those responses.
    Adversary-controlled discovery nodes may send multiple responses to the same \lookup request in order to try to reach that threshold (for example, the adversary may save a SURB from one \lookup request, and use it to send two responses to a later \lookup request; alternatively, the adversary may guess the searcher's username, look up their contact information in the database, and send them Loopix messages directly).
    However, discovery nodes' responses are signed, $U_\challengermath$ knows the public keys of the discovery nodes (which it obtains as part of creating their user account), and we assume the adversary cannot forge signatures from the honest discovery nodes with greater than negligible probability.
    $U_\challengermath$ therefore only accepts one response per \lookup nonce per discovery node signing key.
    Moreover, any malicious mix nodes or active network attackers cannot modify the responses from honest discovery nodes without invalidating the signature.
    Thus, the adversary-generated responses are unable to reach the $f+1$ threshold, and therefore if some SURB and blinded public key are accepted by $U_\challengermath$, they must have been generated by honest discovery nodes.

    \begin{lemma}\label{lemma:message_contents}
    An adversary controlling up to $f$ discovery nodes, and the mix nodes in all but one of the Loopix network layers, gains only a negligible advantage in winning \unlinkabilitygame.
    \end{lemma}

    Let us consider all of the messages that the adversary-controlled nodes receive from the challenger.
    In \lookup, each adversary-controlled discovery node receives a fresh random nonce, the searched username $\mathit{ID}_\challengermath$, and a SURB for replying to $U_b$.
    In \continit, an adversary-controlled discovery node may receive $M_{\mathsf{init}}$, whose mix header is the SURB generated by discovery nodes, and whose payload is $g^a$, the same nonce, and a ciphertext encrypted with the blinded public key of $U_\challengermath$.
    The adversary cannot learn any information from these messages that would increase its advantage in winning the game.
    The Sphinx packet format used by Loopix does not reveal anything about the sender of a message.
    The SURB for $U_b$ is cryptographically indistinguishable from a uniform random bit string \cite{danezis2009sphinx}, and hence does not reveal its destination.
    Knowing that $\mathit{ID}_\challengermath$ is the target of the search does not help distinguish whether $U_0$ or $U_1$ is the searcher, and the nonce and $g^a$ are fresh random values.

    By Lemma \ref{lemma:discovery_response}, the blinded public key $\mathit{bpk}_\mathsf{B}$ that $U_b$ uses to encrypt $M_\mathsf{init}$ is generated by the honest discovery nodes by blinding the public key in the contact information for $U_\challengermath$, and the corresponding private key is not known to \adversary.
    In the case of a lookup of a nonexistent username, $M_{\mathsf{init}}$ is encrypted using the public key in $\Delta_\mathsf{fake}$, which is chosen so that no party knows the corresponding private key, as noted in \S\ref{sec:deterministic-surb-gen}.
    In either case, the ciphertext in $M_{\mathsf{init}}$ cannot be decrypted by the adversary except with negligible probability, and therefore the adversary does not learn the username of $U_b$ or the codeword that $U_b$ included in the message.

    In the case where the adversary controls $U_b$'s provider node or the first mix node along the path of $M_\mathsf{init}$, the adversary would be able to recognise the mix header of $M_\mathsf{init}$ and link it to $U_b$ if $U_b$ were to place it in the network directly.
    However, by reflecting $M_\mathsf{init}$ off a discovery node, with the communication between $U_b$ and the mix node going via Loopix, the adversary becomes unable to link $M_\mathsf{init}$ to any particular sender (even if the reflecting discovery node is controlled by the adversary).

    \begin{lemma}\label{lemma:active_unlinkable}
    An adversary who controls up to $f$ discovery nodes, and who performs an active attack on the \continit and \addfriend protocols, gains only a negligible advantage in winning \unlinkabilitygame.
    \end{lemma}

    Lemma~\ref{lemma:message_contents} already shows that the adversary gains no advantage from the contents of $M_{\mathsf{init}}$.
    The other two messages sent during the \addfriend protocol (from $U_\challengermath$ to $U_b$ in Step \ref{item:addfriend_bob_reply}, and from $U_b$ to $U_\challengermath$ in Step \ref{item:addfriend_alice_reply}) are not normally seen by the adversary (except in encrypted form travelling through the mix network).
    However, through an active attack, the adversary could inject its own SURBs into a protocol run, and thereby \adversary could trick $U_\challengermath$ and $U_b$ into sending these messages to $U_\adversarymath$ instead of each other.

    Intercepting the messages in this way gives the adversary only negligible advantage in winning \unlinkabilitygame.
    Both messages contain only a key-blinded signature and a MAC, and the message from $U_\challengermath$ to $U_b$ additionally contains a Diffie-Hellman key $g^b$.
    The signature is unlinkable, i.e.\ the adversary cannot distinguish between two signatures produced from two separate signing keys, and two signatures produced from the same signing key but with different blinding keys \cite{keyblindingsignatures}; since our protocol generates a different blinding key on every protocol run with overwhelming probability, this means that the signature cannot be linked to any particular user.
    The MAC is based on a key that is not known to \adversary (see Lemma \ref{lemma:active_attack}), and therefore the adversary cannot distinguish it from a random string.
    And finally, $g^b$ is a fresh random value.
    Therefore, even an active attack that redirects messages to the adversary does not impact unlinkability.
    (We examine active attacks on impersonation in Lemma \ref{lemma:active_attack}.)

    \begin{theorem}
    The adversary's advantage over a random guess in winning the unlinkability game \unlinkabilitygame can be reduced to the probability of breaking sender-receiver unlinkability in the underlying Loopix network.
    \end{theorem}

    The adversary gains only negligible advantage from inspecting message contents or returning bad responses at the discovery nodes (Lemma \ref{lemma:message_contents}), or from performing active attacks on the authenticated key exchange protocol (Lemma \ref{lemma:active_unlinkable}).
    This leaves observing or manipulating the traffic between nodes as the only remaining attack, through which the adversary can win \unlinkabilitygame only by breaking sender-receiver unlinkability in the underlying Loopix network (Lemma \ref{lemma:network_traffic}).
  
\subsection{Security against impersonation (G2)}\label{appendix:impersonation_proof}

    \begin{lemma}\label{lemma:active_attack}
    An adversary who controls up to $f$ discovery nodes, and who performs an active attack on the \continit and \addfriend protocols, has only a negligible chance of obtaining the session key $K_s$ in \impersonationgame.
    \end{lemma}

    By controlling at least one discovery node, the adversary learns the username $\mathit{ID}$ that $U_\challengermath$ is searching for, the nonce, the contact information $\Delta_\mathcal{H}$ for user $U_\mathcal{H}$, one or more SURBs that allow \adversary to send messages to $U_\challengermath$ without knowing $U_\challengermath$'s identity, the SURB for constructing $M_\mathsf{init}$, the blinded public key $\mathit{bpk}_\mathsf{B}$ for $U_\mathcal{H}$, and the blinding key $y_\mathsf{B}$ for $U_\mathcal{H}$.
    If $U_\challengermath$ picks an adversary-controlled discovery node as reflector, \adversary additionally learns the Diffie-Hellman key $g^a$ generated by $U_\challengermath$ and the ciphertext in $M_\mathsf{init}$.
    The adversary can use this information to inject modified versions of $M_\mathsf{init}$ and the \addfriend protocol messages (see Figure~\ref{fig:addfriend}) into the network.

    First consider the $M_\mathsf{init}$ message.
    \adversary can construct her own version $M_\mathsf{init}'$ of this message, containing the nonce from $U_\challengermath$.
    If she uses $U_\challengermath$'s DH key $g^a$, her chance of constructing a modified ciphertext that is correctly authenticated under $K_e = \mathsf{KDF}((\mathit{bpk}_\mathsf{B})^a \parallel \text{``init key''})$ is negligible without knowledge of $a$; therefore she must either forward the unmodified ciphertext or replace $g^a$ with a new DH key $g^{a'}$ (where $a' \in \mathbb{Z}_p$) and generate a new ciphertext.
    If she chooses a new DH key, \adversary can include any username and/or codeword of her choice in the new ciphertext (but not the codeword in $M_\mathsf{init}$, which by Lemma \ref{lemma:message_contents} she cannot decrypt), and she can include a SURB that will cause $U_\mathcal{H}$'s reply to be routed back to \adversary.

    $U_\mathcal{H}$ treats the adversary's $M_\mathsf{init}'$ like any other contact request, and decides whether to respond based on the username and/or codeword in the message.
    If $U_\mathcal{H}$ chooses to respond and the SURB in $M_\mathsf{init}'$ is routed to \adversary, the adversary learns $g^b$, a blinded-key signature over the Diffie-Hellman terms, and a MAC over $U_\mathcal{H}$'s identity.
    \adversary can also let the response from $U_\mathcal{H}$ go directly to $U_\challengermath$ without modification.
    However, if \adversary replaced $g^a$ with $g^{a'}$ in $M_\mathsf{init}'$, and the response from $U_\mathcal{H}$ is unmodified, then $U_\challengermath$ will reject the signature, and the active attack on $M_\mathsf{init}$ leads to an abort of the protocol.

    The adversary could also inject an altered version of the message from $U_\mathcal{H}$ to $U_\challengermath$ (\addfriend Step \ref{item:addfriend_bob_reply}) into the network, using one of the SURBs for $U_\challengermath$.
    However, $U_\mathcal{H}$ will only accept a message containing a signature that validates with the blinded public key $\mathit{bpk}_\mathsf{B}$, which by Lemma \ref{lemma:discovery_response} is generated by honest discovery nodes and is not under the adversary's control.
    Due to the existential unforgeability of the key-blinded signature scheme, \adversary has a negligible chance of generating a valid key-blinded signature on this message without knowing $U_\mathcal{H}$'s private key $x_\mathsf{B}$.
    \adversary could replay a signature and $g^b$ Diffie-Hellman term from another protocol run, but in that case \adversary would not know $b$, hence \adversary cannot compute $K_m$, and hence \adversary has a negligible chance of forging the MAC on this message.
    In any case, the adversary cannot inject an altered version of this message that is accepted by $U_\challengermath$.

    Finally, the adversary could inject an altered version of the final message from $U_\challengermath$ to $U_\mathcal{H}$ (\addfriend Step \ref{item:addfriend_alice_reply}).
    If $M_\mathsf{init}$ was previously unaltered, $U_\mathcal{H}$ will accept this message only with a key-blinded signature that is validated with $U_\challengermath$'s blinded public key $\mathit{bpk}_\mathsf{A}$ (which was either included unaltered in $M_\mathsf{init}$, or which was obtained from $U_\challengermath$'s username in $M_\mathsf{init}$ via \lookup, which by Lemma \ref{lemma:discovery_response} is generated by honest discovery nodes and is not under the adversary's control); by a similar argument to the previous paragraph, \adversary cannot construct a modified message where both the signature and the MAC are valid.

    If $M_\mathsf{init}$ was previously replaced by $M_\mathsf{init}'$, then the last message in the \addfriend protocol must have also been generated by \adversary, since $U_\challengermath$ would abort the protocol and would not generate such a message.
    In this case, since the username or blinded public key and the Diffie-Hellman term $g^{a'}$ in $M_\mathsf{init}'$ are under the adversary's control, \adversary is able to produce a valid key-blinded signature and MAC for this message.
    However, the session key derived from this protocol run is $K_s' = \mathsf{KDF}(g^{a'b} \parallel \text{``session key''})$, whereas $U_\challengermath$ computed the session key as $K_s = \mathsf{KDF}(g^{ab} \parallel \text{``session key''})$.
    $K_s$ and $K_s'$ are different with overwhelming probability, and \adversary cannot compute $K_s$ since she does not know $a$.

    \begin{theorem}
    The adversary cannot win the impersonation game \impersonationgame with a significantly better chance than random guessing.
    \end{theorem}

    By Lemma \ref{lemma:active_attack}, the adversary does not obtain the session key $K_s$.
    Therefore, when $U_\challengermath$ encrypts $m_b$ with $K_s$ using a symmetric cipher that provides indistinguishability under chosen plaintext attack, \adversary has only negligible advantage over randomly guessing $b$.    
\newpage\section*{Meta-review}
\subsection*{Summary of Paper}

This paper proposes Pudding, a user-discovery mechanism for mix networks. User names are email addresses, and users register their public key and (mixnet) location information with discovery nodes, verified by email. When wishing to contact another user, the users reach out to the discovery nodes to fetch a single-use reply block for the destination user, along with that user's blinded public key. Pudding provides user discovery with unlinkability, security against impersonation, external identity verification, and membership unobservability. The authors implement Pudding in Rust on top of the Nym network and show that it is useful in practice.

\subsection*{Scientific Contributions}

\begin{itemize}
    \item Creates a New Tool to Enable Future Science
    \item Provides a Valuable Step Forward in an Established Field
\end{itemize}

\subsection*{Reasons for Acceptance}

\begin{itemize}
    \item This paper provides a valuable step forward in an established field. User discovery is an important aspect of, but is often missing from, anonymous communication systems. Pudding provides this functionality for Loopix, a prominent mixnet design.
    \item This paper creates a new tool to enable future science. The authors wrote an implementation of Pudding in Rust that works with the Nym anonymity network that the authors will release open-source under an MIT license.
\end{itemize}

\subsection*{Noteworthy Concerns}

Much trust is placed in the Pudding discovery nodes. These nodes:

\begin{itemize}
    \item Know who is registered in the system
    \item Know the location (i.e., provider identity) of each registered user
    \item Know when every user query is made
\end{itemize}

As a result, membership unobservability is provided only against adversary-controlled clients but not servers. This strong trust assumption may not be tolerable in practice; future work should consider ways to relax the trust assumption placed on these nodes.

\end{document}